\documentclass[twocolumn]{aastex701}
\usepackage{apjfonts}
\usepackage{subfigure}
\usepackage{multirow}
\usepackage{amsmath}
\usepackage{booktabs}
\usepackage{threeparttable}
\usepackage[shortlabels]{enumitem}

\begin{document}

\title{Insights from the ``Red devil" AT 2022fpx: A Dust-reddened Family of Tidal Disruption Events Excluded by Their Apparent Red Color?}

\author[orcid=0000-0003-4959-1625,sname='Lin',gname='Zheyu']{Zheyu Lin}
\affiliation{Department of Astronomy, University of Science and Technology of China, Hefei 230026, China}
\affiliation{School of Astronomy and Space Sciences, University of Science and Technology of China, Hefei, 230026, China}
\email[show]{linzheyu@mail.ustc.edu.cn}

\author[0000-0002-7152-3621]{Ning Jiang}
\affiliation{Department of Astronomy, University of Science and Technology of China, Hefei 230026, China}
\affiliation{School of Astronomy and Space Sciences, University of Science and Technology of China, Hefei, 230026, China}
\affiliation{Frontiers Science Center for Planetary Exploration and Emerging Technologies, University of Science and Technology of China, Hefei, Anhui, 230026, China}
\email[show]{jnac@ustc.edu.cn}

\author[0000-0003-4225-5442]{Yibo Wang}
\affiliation{Department of Astronomy, University of Science and Technology of China, Hefei 230026, China}
\affiliation{School of Astronomy and Space Sciences, University of Science and Technology of China, Hefei, 230026, China}
\email{wybustc@ustc.edu.cn}

\author[0000-0002-7660-2273]{Xu Kong}
\affiliation{Department of Astronomy, University of Science and Technology of China, Hefei 230026, China}
\affiliation{School of Astronomy and Space Sciences, University of Science and Technology of China, Hefei, 230026, China}
\affiliation{Frontiers Science Center for Planetary Exploration and Emerging Technologies, University of Science and Technology of China, Hefei, Anhui, 230026, China}
\email[show]{xkong@ustc.edu.cn}

\author[0000-0001-7689-6382]{Shifeng Huang}
\affiliation{Department of Astronomy, University of Science and Technology of China, Hefei 230026, China}
\affiliation{School of Astronomy and Space Sciences, University of Science and Technology of China, Hefei, 230026, China}
\email{sfhuang999@ustc.edu.cn}

\author[0000-0001-8078-3428]{Zesen Lin}
\affiliation{Department of Physics, The Chinese University of Hong Kong, Shatin, N.T., Hong Kong S.A.R., China}
\email{zesenlin@ustc.edu.cn}

\author[0009-0000-0126-8701]{Chen Qin}
\affiliation{Department of Astronomy, University of Science and Technology of China, Hefei 230026, China}
\affiliation{School of Astronomy and Space Sciences, University of Science and Technology of China, Hefei, 230026, China}
\email{qinc@mail.ustc.edu.cn}

\author[0000-0002-8438-8529]{Tianyu Xia}
\affiliation{Department of Astronomy, University of Science and Technology of China, Hefei 230026, China}
\affiliation{School of Astronomy and Space Sciences, University of Science and Technology of China, Hefei, 230026, China}
\email{xiatianyu@mail.ustc.edu.cn}

\begin{abstract}
\noindent We report unnoticed but intriguing features in the peculiar nuclear transient AT 2022fpx, and investigate its type. These features include the constantly red optical color of $g-r>0$, stable soft X-ray flare ($kT\sim100$ eV) in the past $\sim$550 days, a prominent mid-infrared echo peaked at $\sim$$10^{43.3}$ erg s$^{-1}$ and the confirmation of a weak active galactic nucleus by weak flares in pre-event Wide-field Infrared Survey Explorer mid-infrared light curves with no contemporary optical, radio or X-ray counterparts. The combination of the optical red color and possible origin of a tidal disruption event (TDE) of AT 2022fpx is particularly attractive, as it challenges the most widely accepted and adopted ``blue color'' criterion for optical TDE selection. Although we still cannot confirm whether the red color is intrinsic, we do find that the  ``blue color'' criterion can filter out normal TDEs whose optical-UV spectral energy distributions (SEDs) are either severely contaminated by prominent emission lines (especially H$\alpha$) or heavily dust-reddened. Hence, its potential selection effect may have been imprinted on the whole optical TDE family. Blackbody fitting on the optical (rest-frame $\sim$$4000-7000$ \AA) and optical-UV ($\sim$$2000-7000$ \AA) SEDs of four TDEs with high-cadence UV observations shows that $T_\mathrm{bb}$ rise by $\sim$40$-$110 \% when the UV bands are included. The power-law models ($f_{\lambda}\propto\lambda^{-\alpha}$ with $\alpha=2-3$) can fit the rest-frame $\sim$$2000-7000$ \AA\ SEDs more consistently, indicating that SEDs should peak at shorter wavelengths, but not simple blackbodies. Hence, the estimated released energy for the optical-UV bright but X-ray faint TDEs based on blackbody SED fitting should be significantly lower than the intrinsic energy.

\end{abstract}



\section{Introduction} \label{sec:intro}

A tidal disruption event (TDE) usually happens when an unlucky star wanders too close to a black hole (BH) in the center of a galaxy, that its self-gravity is overcome by the tidal forces of the BH. Although the first TDE was discovered in the X-ray band \citep{Bade1996,Komossa1999}, the wide-field and high-cadence time-domain surveys in optical bands have dominated the discovery of TDEs in the past decade, making the number of TDE to $\sim$100. The dominance of optical bands should continue, as the recently commissioning Wide Field Survey Telescope \citep[WFST;][]{WFST} and the forthcoming Vera Rubin Observatory \citep[VRO;][]{Ivezic2019} are predicted to be capable of discovering hundreds to thousands of TDEs \citep[e.g.,][]{Strubbe2009,vanVelzen2011,Thorp2019,Bricman2020,Roth2021,Lin2022a,Wang2023}.

A TDE in X-ray is characterized by two features \citep[e.g.,][]{Komossa2015,Saxton2020}. First, an outburst appears in the center of a galaxy, whose spectrum can be well described by a soft blackbody with $kT\sim50-100$ eV, followed by a power-law like declining phase, in which the luminosity fades for more than one order of magnitude. Second, its host galaxy is quiescent in optical, radio and X-ray bands before this outburst.

A TDE in optical and UV bands is mainly characterized by three features. First, the spectral energy distribution (SED) can be well described by a blackbody component, with the temperature $T_{\rm bb}\sim(1-5)\times10^4$ K that varies slowly. This results in the steady blue color in optical bands (e.g., $g-r<0$). Second, the optical light curve of a TDE usually shows a monthly rise to a peak blackbody luminosity of $L_{\rm{bb}}\sim10^{43-46}$ erg s$^{-1}$ \citep{Lin2022b,Hammerstein2023,Yao2023}, followed by a power-law like decline that lasts for months to years. Third, a blue continuum is always shown in the spectrum of a TDE. Meanwhile, the broad H$\alpha$, H$\beta$ or He $\textsc{ii}$ emission lines with full wavelength half maxima (FWHMs) of $\sim$10000 km s$^{-1}$ usually appear near the peak \citep{vanVelzen2021}, and then gradually narrow and weaken as the luminosity declines. Nonetheless, the blue continuum may dominate the spectrum, which results in a featureless spectrum \citep{Hammerstein2023}. 

In real transient surveys, in order to perform the follow-up observations, TDE candidates should be selected quickly. Apart from the steady blue color and the shape of the light curves, several additional features, e.g., the pre-flare variability, the offset from the host center, mid-infrared (MIR) colors \citep{Stern2012,vanVelzen2021}, can effectively exclude the impostors
(See \citealt{Zabludoff2021} for a review). Extensive 
follow-up multiwavelength observations and spectroscopy are still necessary for an accurate classification, since
some impostors manage to escape from the careful photometric filtering, including some
supernovae (SNe) that do not undergo significant color evolution \citep{Nyholm2020},
outbursts in narrow-line Seyfert 1 (NLSy1) galaxies \citep{Frederick2021}, and the so-called ``turn-on active galactic nuclei (AGNs)'', in which accretion rate suddenly rises, producing a turn-on flare that changes a quiescent galaxy into a broad-line AGN within months to years \citep{Gezari2017, Sanchez2024}.



The steady blue color is probably the most widely accepted TDE feature and frequently used selection criterion. However, it was initially not a predicted feature, but a surprising similar feature in the first two optical TDEs that were systematically found in the Stripe 82 data (see Fig. 12 of \citealt{vanVelzen2011}). Thus, the ``color'', by its nature, is not a robust selection criterion and even not yet well explained. In other words, given other robust observational evidence, TDEs can show different color features.

In this paper, we focus on a peculiar nuclear transient, the ``red devil'' AT 2022fpx (also known as ATLAS22kjn, ZTF22aadesap and Gaia22cwy). It was initially spectroscopically classified as a possible H+He TDE at $z=0.073$ by \citet{Perez-Fournon2022} during its rise stage, as strong H$\alpha$, H$\beta$, He \textsc{ii} $\lambda$4686, and H$\gamma$ emission lines exhibited in its spectrum. Its ZTF color and continuum was noted as redder than typical TDEs, probably due to the obscuration in the nucleus of the host galaxy. After the classification, the Swift X-Ray Telescope \citep[XRT;][]{Burrows2005} and Ultra-Violet/Optical Telescope \citep[UVOT;][]{Roming2005} were triggered (PI: Guolo). We were intrigued by the TDE classification, as the red color ($g-r\sim$ 0.4) is against the ``blue color'' optical TDE selection criterion, and proposed for further Swift observations to capture its later UV evolution and possible delayed X-ray flare. Meanwhile, we carried out extensive follow-up spectroscopy in order to investigate the evolution of its emission lines. 

During the preparation of this work, \citet{Koljonen2024} published their work on this transient, in which they well studied its prominent iron coronal lines, variable polarization, delayed X-ray flare and tentatively interpreted this transient as a long-lived TDE. In this article, we focus on characteristics that remain unnoticed, including the recent X-ray plateau of $\sim$550 days that is unusual for X-ray TDEs, the historical weak but genuine MIR variability that favors a weak AGN, and most importantly, the red optical color.

The paper is organized as follows. In Section \ref{sec:reduce}, the procedures of observations and the data reduction are introduced. In Section \ref{sec:analyze}, we analyze the host galaxy, the historical and recent photometric evolution in UV, optical and MIR bands, and the X-ray and optical spectra. In Section \ref{sec:discuss}, we discuss the possible origins of AT 2022fpx. We reach a final summary in Section \ref{sec:conclusion}. All errors marked by ``$\pm$'' represent the 1$\sigma$ confidence intervals. We adopt a flat cosmology with $H_{0} =70$ km~s$^{-1}$~Mpc$^{-1}$ and $\Omega_{\Lambda} = 0.7$. Based on six spectra, we adopt the median redshift $z=0.0735$ for AT 2022fpx. We use the extinction law from \citet{Fitzpatrick1999}
with $R_V=A_V/ E(B-V)=3.1$ \citep{Osterbrock2006} and adopt a Galactic extinction of $E(B-V) = 0.0177$ mag \citep{Planck2016}. All magnitudes are in the AB \citep{Oke1974} system.

\section{Observation \& Data Reduction} \label{sec:reduce}

\subsection{ZTF \& ATLAS Optical Photometry} \label{sec:ztfatlas}
\begin{figure*}[htb!]
    \centering
    \includegraphics[scale=0.35]{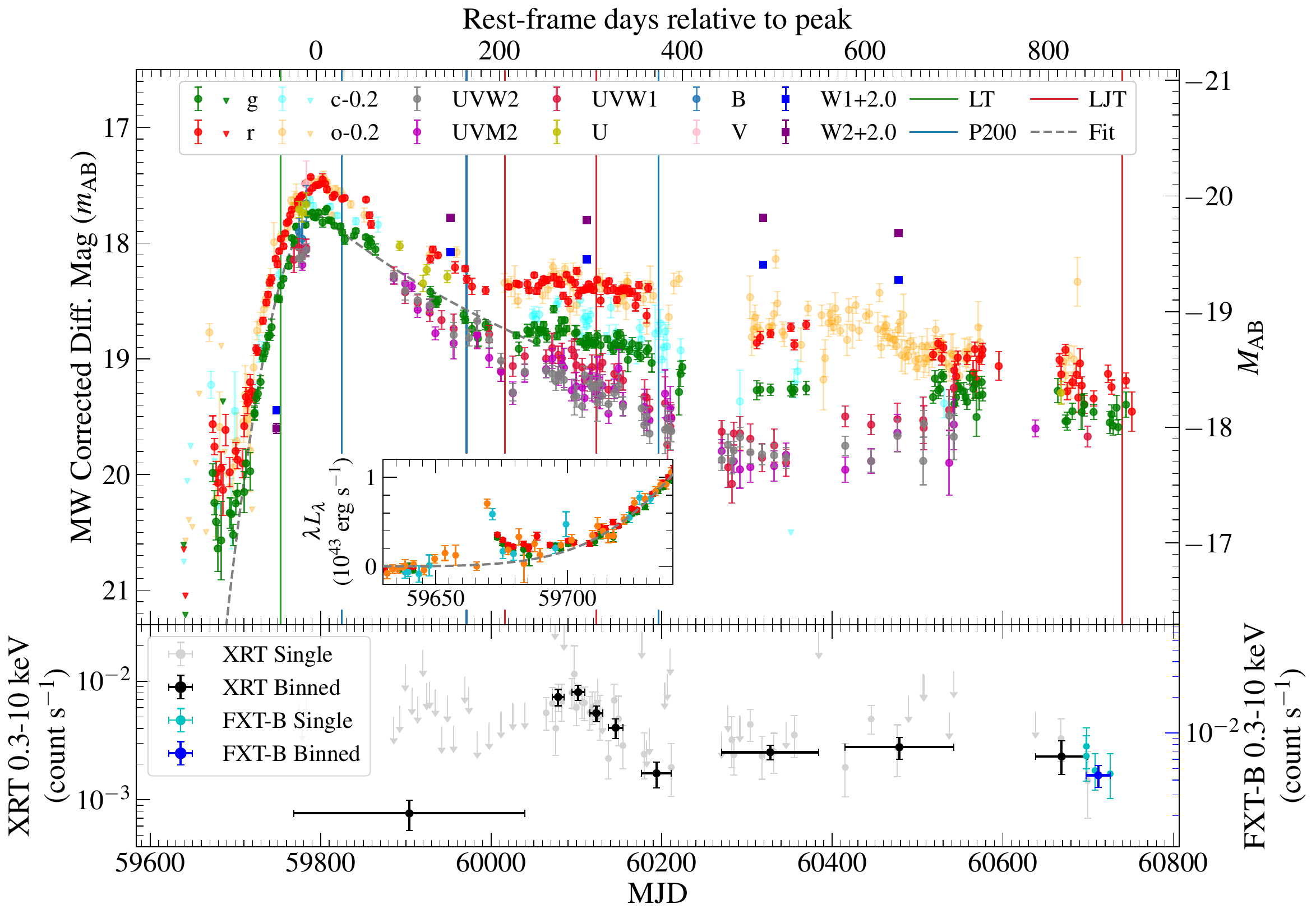}
    \caption{\textbf{Top panel:} The Galactic-extinction-corrected differential light curves of AT 2022fpx. For non-detections, 3$\sigma$ upper limits are plotted in down triangles. The dates that spectra were taken are shown in vertical lines. The gray dashed line shows the best fit of $g$-band light curve to the rise-decline function (Equation \ref{eqn:fitg}). The inset is a zoom-in view of the very early stage in which a spiky precursor can be clearly revealed. \textbf{Bottom panel:} The Swift XRT and EP FXT-B light curve of observed $0.3-10.0$ keV count rates. For each single XRT epoch, the 3$\sigma$ upper limit for each non-detection is plotted in a light gray down arrow, while for each detection, the photon count rate and error are plotted in a light gray dot and error bars. For the first 23 XRT epochs, the X-ray emission did not reach the detection threshold in any single epoch, while a binned event with a total exposure time of 33.1 ks reveals a detection with a low count rate. A rapid enhancement in X-ray bands then result in detections in single epochs. After that, the luminosity declined and has maintained at a higher level than the earliest state, so we bin the neighboring observations after the enhancement to reach a higher SNR. The binned measurements are plotted in black. The count rates of very recent FXT-B epochs are plotted in cyan, we bin these epochs into one bin, which is displayed in blue. The y-axis is normalized by the effective area ratio (300 cm$^2$ vs. 110 cm$^2$).}
    \label{fig:lc}
\end{figure*}

The ZTF differential point-spread-function (PSF) photometry is obtained through the ZTF Forced-Photometry Service \citep{Masci2019}. We perform the filtering and the baseline correction, then build the $g$- and $r$- band light curves. We also obtain the ATLAS differential photometry from the ATLAS forced photometry server \citep{Shingles2021}, and build the 1-day binned $c$- and $o$- band light curves. The Galactic extinction corrected light curves are shown in the top panel of Figure \ref{fig:lc}.

\subsection{Swift} \label{sec:swift}
Until March 1, 2025, 75 observations have been performed by Swift XRT and UVOT under a series of target-of-opportunity (ToO) requests (ToO ID: 17570, PI: Guolo; ToO ID: 18061, PI: Jiang; ToO ID: 18079, PI: Koljonen; ToO ID: 18347/18743/18781/18880/18963/19068/19100/19252/19352/\\19706/19846/20091/20612/21495, PI: Lin). We retrieve the Swift data from HEASARC\footnote{\url{https://heasarc.gsfc.nasa.gov/cgi-bin/W3Browse/swift.pl}}, and process all data with \texttt{heasoft} v6.33. The reduction procedures are described below.

\subsubsection{UVOT} \label{sec:uvot}
For each UVOT image file, we exclude the extensions with bad photometric flags, and use \texttt{uvotimsum} to sum image files with multiple valid extensions. Then, we define the source and background regions as a circle of the radius of $7^{\prime\prime}$ and $20^{\prime\prime}$, respectively. After that, the task \texttt{uvotsource} performs photometry on each image. To get the differential magnitudes in UVOT bands, we calculate the reference magnitudes of the host galaxy by fitting its spectral energy distribution (SED, see Section \ref{sec:sed} for details), and subtract them from the measured magnitudes. In Figure \ref{fig:sed}, the synthetic AB magnitudes are plotted on the best-fit SED.

\subsubsection{XRT} \label{sec:xrt}
For each XRT epoch, we reduce the data by \texttt{xrtpipeline} and obtain the level 2 products. Then we use \texttt{xrtproducts} to extract the level 3 products. The source region is selected as a circle of the radius of $20^{\prime\prime}$, while the background region is defined as a source-free annulus with an inner radius of $80^{\prime\prime}$ and an outer radius of $200^{\prime\prime}$. We obtain the source and background photon counts in 0.3-10 keV by \texttt{ximage}. For images with photon counts in source region $N\leqslant15$, a Bayesian approach is applied to calculate the 3$\sigma$ lower and upper limits \citep{Kraft1991}; While for $N>15$, a Gaussian approach is adopted \citep{Evans2007,Evans2009}. The X-ray light curve is displayed in the bottom panel of Figure \ref{fig:lc}.

\subsection{Einstein Probe (EP) X-ray Observations} \label{sec:ep}
AT 2022fpx has been accepted as one of the EP Cycle 1 targets under our proposal titled ``Monitoring of Extreme Coronal Line Emitters" (Proposal No.: Cycle1-0046). Observations were conducted by the Follow-up X-ray Telescope (FXT) from January 22, 2025 to February 19, 2025. \texttt{fxtsoft} v1.10 is employed to reduce all FXT data. Following the recommendation of the EP team, we discard the FXT-A data which are currently not reliable in flux calibration, and only use the reliable FXT-B data. Combining all epochs yields a total exposure time of 6.9 ks. The source region is selected as a circle of the radius of $40^{\prime\prime}$, while the background region is defined as a source-free annulus with an inner radius of $80^{\prime\prime}$ and an outer radius of $200^{\prime\prime}$. The calculation of the count rate and the extraction of the spectrum are similar as the procedure of XRT. The FXT-B light curve is displayed in the bottom panel of Figure \ref{fig:lc}.

\subsection{WISE MIR Photometry} \label{sec:wise}
AT 2022fpx has been continuously observed by the Wide-field Infrared Survey Explorer \citep[WISE;][]{Wright2010}, and the successive Near Earth Object Wide-field Infrared Survey Explorer \citep[NEOWISE;][]{Mainzer2011,Mainzer2014}, at W1 (3.4 $\mu$m) and W2 (4.6 $\mu$m) bands every half year. 

To study the possible MIR dust echo, we first query and download the W1- and W2-band photometric data from the AllWISE Multiepoch Photometry Table and NEOWISE-R Single Exposure (L1b) Source Table. We filter out the bad data points that have 
\texttt{NaN} magnitudes and errors; or get affected by 
a nearby image artifact (\texttt{cc\_flags} $\neq$ 0), the scattered moon light (\texttt{moon\_masked} $\neq$ 0), or a nearby detection (\texttt{nb} $>$ 1). The remaining data points are binned in $\sim$half-year bins to improve the signal-to-noise ratio (SNR). 

To obtain more accurate measurements, we attempted to perform PSF photometry on the difference images of the time-resolved WISE/NEOWISE coadds~\citep{Meisner2018} with the first epoch as a reference following \citet{Jiang2021}. Variability is clearly detected in both bands ($\gtrsim$5$\sigma$) at two epochs before 2022, i.e., MJD = 56667 and MJD = 57396, yet much weaker than the post-outburst echo. It is strongly suggestive of weak AGN activity in this galaxy.

\subsection{Archival CSS \& PTF Photometry}\label{sec:arch}
To find out if any variability had happened before this outburst, we query the archival Catalina Sky Survey (CSS) and Palomar Transient Facility (PTF) catalogs and combine the data into 30-day bins to improve the SNR. The results are displayed and discussed in Section \ref{sec:preflare}.

\subsection{Optical Spectroscopy} \label{sec:spec}
Apart from the spectrum that was taken by the SPectrograph for the Rapid Acquisition of Transients (SPRAT) on the Liverpool Telescope (LT) and submitted to TNS \citep{Perez-Fournon2022}, we also obtained 3 spectra using the Double Spectrograph \citep[DBSP;][]{Oke1982} on the 200 inch Hale telescope at the Palomar Observatory (P200), and 3 spectra using the Yunnan-Faint Object Spectrograph and Camera (YFOSC) on the Lijiang 2.4m telescope (LJT). The observation dates for these spectra are labeled in vertical lines on the top panel of Figure \ref{fig:lc}. We use \texttt{pypeit} to reduce the P200/DBSP spectra, and extract the LJT/YFOSC spectra by \texttt{IRAF}.

\subsection{Radio Observations}
The position of AT 2022fpx has been observed by the Very Large Array Sky Survey \citep[VLASS;][]{Lacy2020} for three times. Two of the observations were performed before the flare, while the other was taken $\sim$300 days after the outburst. In all three epochs, no source has been detected within a radius of 1$^\prime$. Therefore, we would not discuss the radio properties in the following texts.

\section{Data Analysis} \label{sec:analyze}
\subsection{SED Fitting of the Host Galaxy} \label{sec:sed}
As mentioned in Section \ref{sec:arch}, no host galaxy spectrum was taken before the flare, and as the flare is still lasting, it is not realistic to obtain the host galaxy spectrum now. Therefore, we build its SED using photometry from the Sloan Digital Sky Survey DR17 \citep[SDSS;][]{SDSSDR17}, the Panoramic Survey Telescope and Rapid Response System \citep[Pan-STARRS;][]{Flewelling2020,Waters2020} DR2, the Dark Energy Spectroscopic Instrument (DESI) Legacy Survey DR10 \citep{Dey2019}, the Two Micron All-Sky Survey \citep[2MASS;][]{Skrutskie2006}; the unWISE catalog \citep{Schlafly2019}, and AllWISE catalog \citep{Cutri2014}. After correcting for the Galactic extinction, an SED fitting is performed on the Code Investigating GALaxy Emission \citep[\texttt{CIGALE}, v2025.0;][]{Boquien2019}. We adopt an exponential star formation history (SFH) with an optional exponential burst, a single stellar population of \citet{Bruzual2003}, a modified \citet{Calzetti2000} dust attenuation law, a dust emission component with contribution of AGN component that is parameterized by \citet{Dale2014}.

The SED can be well fitted (with reduced $\chi^2$ of 0.31) with weak contribution of the AGN component ($f_{\rm AGN}=0.15_{-0.07}^{+0.09}$). The stellar mass derived from the Bayesian approach is $M_*=6.89_{-1.93}^{+2.95}\times10^9\,M_\odot$. Using the relation between the black hole and stellar mass of \citet{Reines2015}, we derive a black hole mass of log ($M_{\rm BH}/M_\odot$) = $6.24 ^{+0.32}_{-0.33}$. 

\begin{figure*}[htb!]
    \centering
    \centering
    \subfigure[\texttt{CIGALE} best-fit model]{
        \includegraphics[width=8cm]{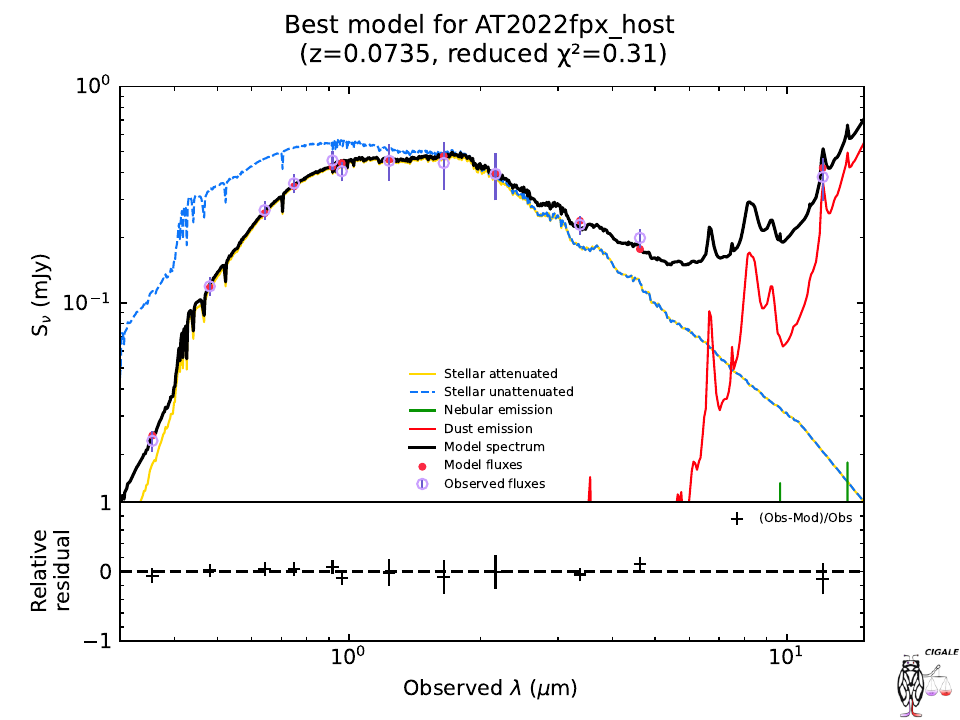}
    }
    \subfigure[Swift UVOT synthetic magnitudes]{
        \includegraphics[width=8cm]{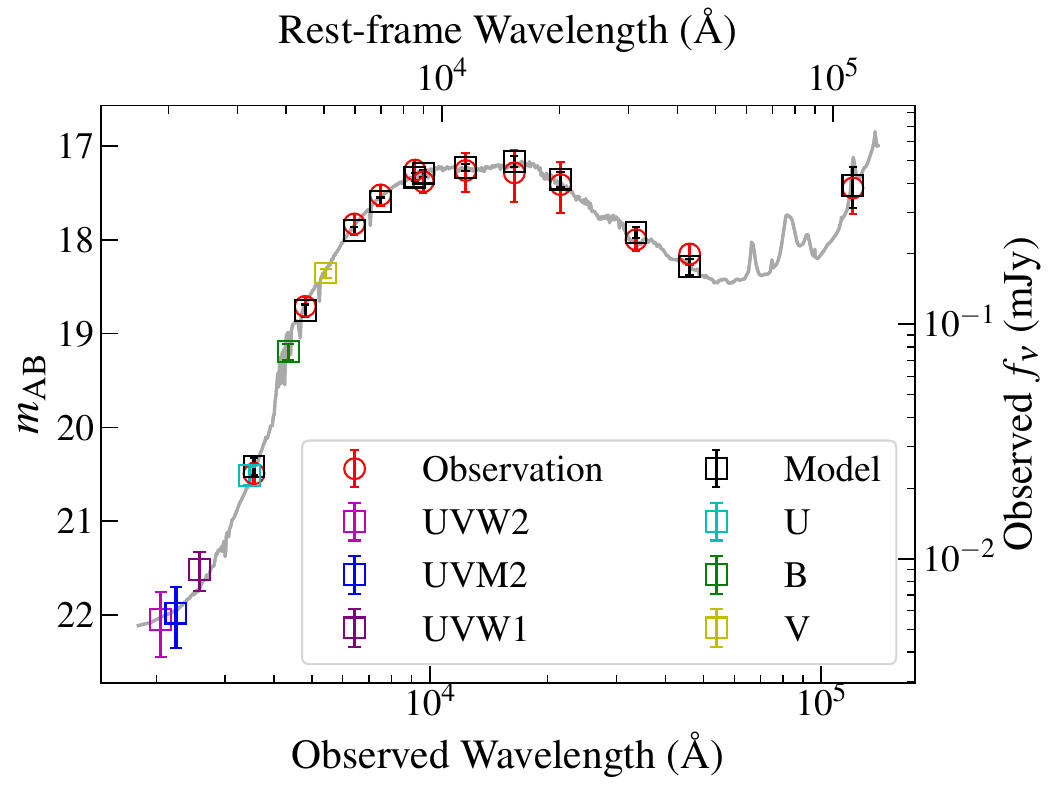}
    }
    \caption{\textbf{Left panel:} The best-fit \texttt{CIGALE} model on the host-galaxy SED yields a perfect match with reduced $\chi^2=0.31$. \textbf{Right panel:} The synthetic magnitudes of six UVOT bands (UVW2, UVM2, UVW1, U, B, V).}
    \label{fig:sed}
\end{figure*}

The best fit is shown in AB magnitudes and millijanskys (mJy) in Figure \ref{fig:sed}, while the best optical spectrum is plotted at the bottom of Figure \ref{fig:spec}.

\subsection{UV, Optical \& MIR Photometric Analysis}
\subsubsection{Before the Outburst}\label{sec:preflare}
\begin{figure*}[htb!]
    \centering
    \includegraphics[scale=0.5]{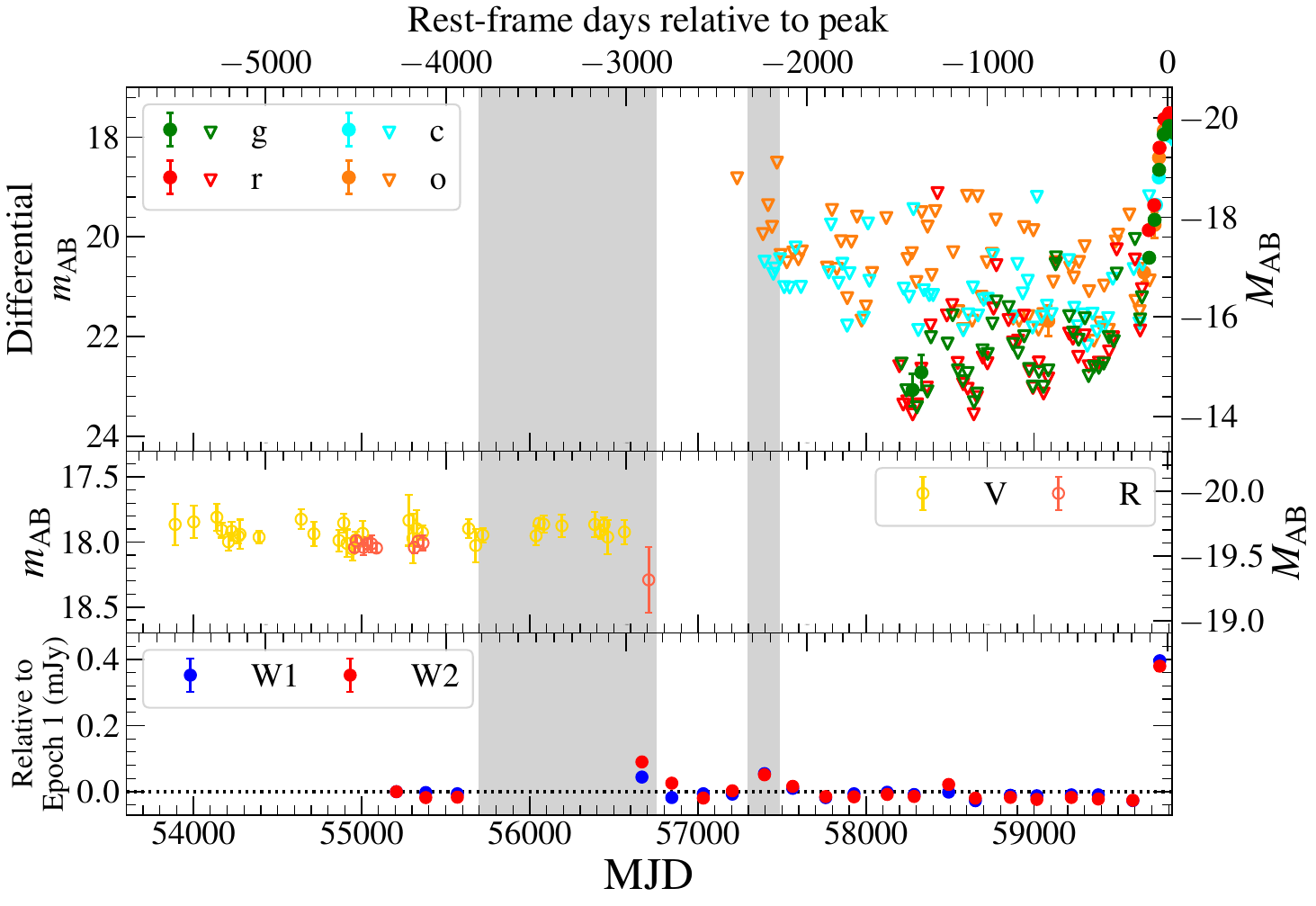}
    \caption{The pre-peak light curves of the position of AT 2022fpx. \textbf{Top and middle panel:} Differential $g$-, $r$-, $c$- and $o$-band light curves and non-subtracted V- and R-band light curves. To improve the SNR, for each band, we combine the data into 30-day bins; \textbf{Bottom panel:} MIR bands W1 and W2. We utilize the unWISE stacked images of each epoch, employ PSF photometry after subtracting the image of the first epoch.
    Two small flares are shown but no optical counterparts can be found at those periods (marked in gray).}
    \label{fig:preflare}
\end{figure*}

 In Figure \ref{fig:preflare}, we plot the pre-peak host-subtracted light curves in these bands: ATLAS $c$ and $o$ bands, ZTF $g$ and $r$ bands, WISE W1 and W2 bands. Also, the non-subtracted CSS V-band and PTF R-band light curves are displayed. The figure indicates no significant variability before the outburst. Based on unWISE stacked images, its pre-outburst MIR color is W1 $-$ W2 = 0.49~$\pm$~0.02, which is below the AGN threshold of W1$-$W2 $>$ 0.8 \citep{Stern2012}. Despite this, as discussed in Section \ref{sec:wise}, 
 PSF photometry on the subtracted images reveals that both the W1 and W2 bands show weak but clear enhancement in two early epochs of the NEOWISE project, indicating AGN activity inside, while its intensity is constrained by the lack of optical counterpart.
 

\subsubsection{Since the Outburst}\label{sec:flare}
The outburst started with a spiky precursor in the optical bands (see the inset of Figure \ref{fig:lc} for a clear view). It was first detected on MJD = 59669.5 in the ATLAS $o$ band with $m_o=18.97\pm0.08$ (binned). The last non-detection was only 4 days before that (MJD = 59665.5), with a tight 3$\sigma$ upper limit of $m_o=20.70$. It then faded and stayed at $m_o\sim20$ for about a month. After that, its optical luminosity steadily rose and reached the peak at MJD $\sim$ 59795 ($t_{\rm peak}$). After the peak, its optical/UV luminosity declined and has turned into a slower decline since MJD $\sim$ 60050 ($t_{\rm turn}$). To evaluate the rate of rise and decline in luminosity, we mask the precursor and fit the $g$-band light curve before this turn with a rise-decline function:
\begin{align}\label{eqn:fitg}
    f_g(t)
    &=f_g(t_{\rm peak})\times
    \begin{cases}
        {\rm exp}[{-(t-t_{\rm peak})^2/(2\sigma^2)}], & t\leqslant t_{\rm peak} \\
        \left(\frac{t-t_{\rm peak}+\tau}{\tau}\right)^{-5/3},  & t_{\rm peak}<t\leqslant t_{\rm turn}.
    \end{cases}
\end{align}

The best-fit parameters are $\sigma=41.5\pm0.7$ d, $\tau=296\pm9$~d. From these parameters we derive the rest-frame duration from the half-max of the luminosity to the max ($t_{\rm 1/2,rise}$) and from the max to the half-max ($t_{\rm 1/2,decline}$) as $t_{\rm 1/2,rise}=48.9\pm0.8$~d, $t_{\rm 1/2,decline}=152.7\pm4.6$~d and their sum $t_{\rm 1/2}\sim200$~d. We will compare these timescales with those of other nuclear transients in Section \ref{sec:optuvevo}. As seen in the top panel of Figure \ref{fig:lc}, the best-fit curve (gray line) well describes the $g$-band light curve before the plateau.

In addition, we fit the UV and optical light curves into a blackbody SED by the \texttt{Superbol} package \citep{Nicholl2018}. It uses the light curve of one band as a reference and interpolate the sparser sampled light curves. For each UVOT epoch, we fit the three well sampled UV bands UVW2, UVM2, UVW1 and optional U band together with the optical bands $g$, $r$, $c$ and $o$. The reference band is selected as UVW2. For comparison, we also fit only optical bands with the reference of $g$-band light curve.

\begin{figure*}[htb!]
    \centering
    \subfigure[Blackbody fitting results]{
    	\includegraphics[width=8cm]{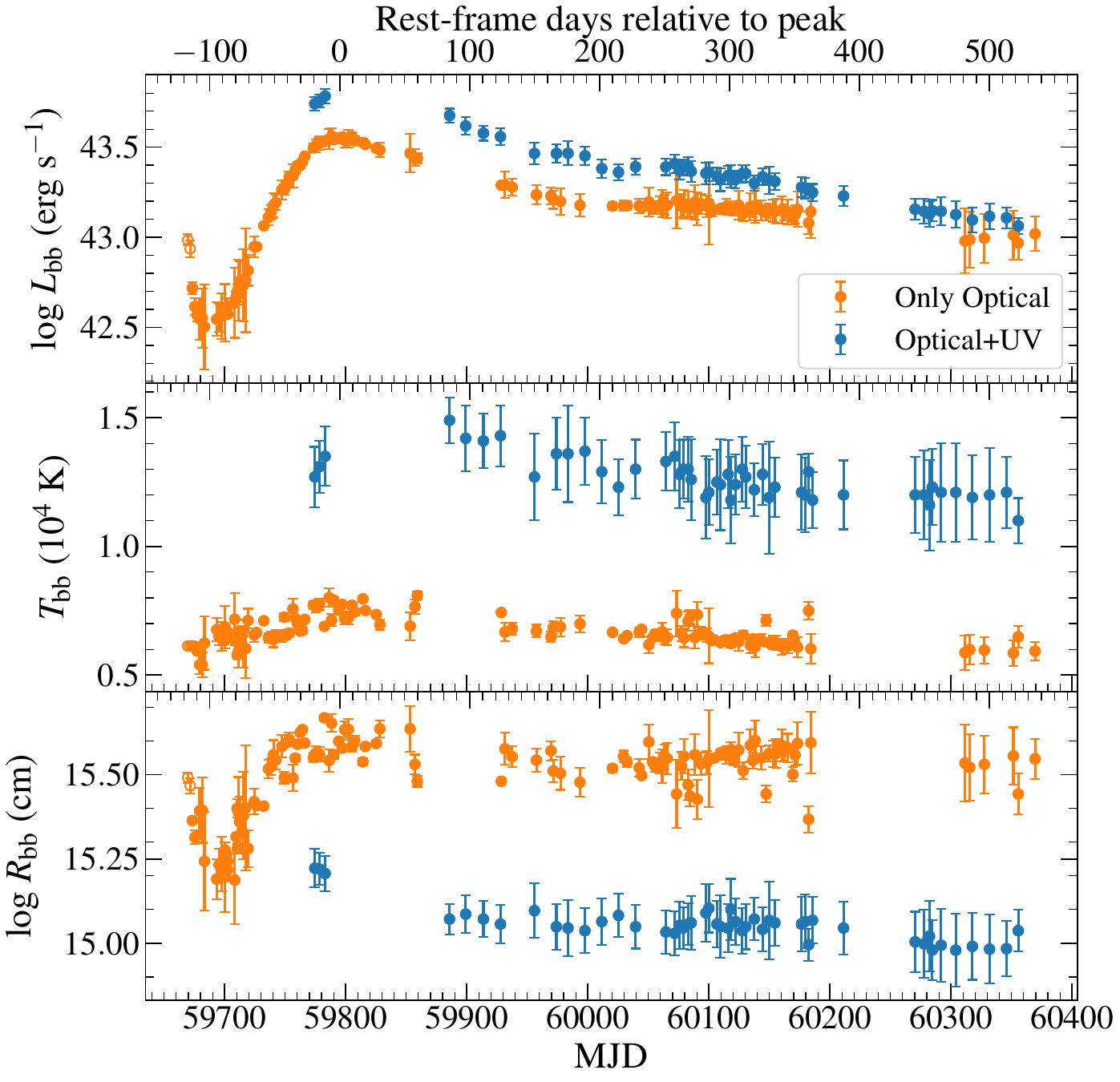}
    }
    \centering
    \subfigure[Power-law fitting results]{
    	\includegraphics[width=8cm]{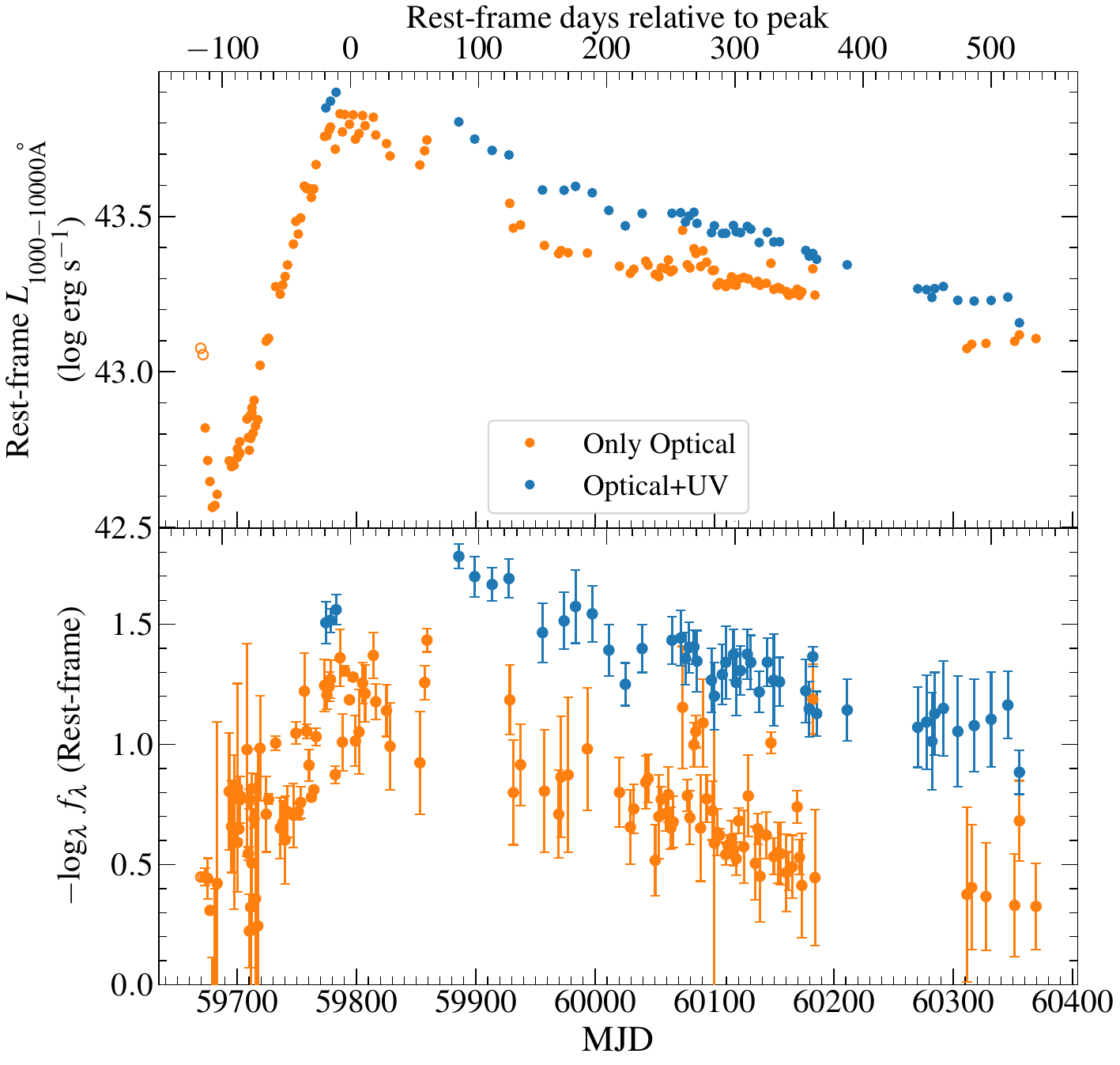}
    }
    \caption{The fitting results for UV and optical photometric data. The optical bands are $g$, $r$, $c$, $o$, while UV bands are UVW2, UVM2, UVW1, U. The details for the fitting are introduced in Section \ref{sec:flare}. \textbf{Left:} The blackbody luminosity $L_{\rm bb}$, temperature $T_{\rm bb}$ and radius $R_{\rm bb}$. Since the first two epochs only have one band photometry, their blackbody temperatures are fixed as the best-fit temperature of the third epoch, so as the power-law indexes. \textbf{Right:} The rest-frame 1000$-$10000 \AA\ luminosity and the power-law index $\alpha$ ($f_{\lambda}\propto\lambda^{-\alpha}$, rest-frame).}
    \label{fig:bb+po}
\end{figure*}

As shown in the left panel of Figure \ref{fig:bb+po}, the involvement of three UV bands significantly rises the fitted blackbody temperature from $T_{\rm bb}\sim7500$ K to $\sim14000$ K. This result confirms a UV excess towards a red blackbody model.

We instead fit the same light curves into a power-law model ($f_{\lambda}\propto\lambda^{-\alpha}$, rest-frame), and find no significant improvement, as the involvement of three UV bands greatly lift the slope. The result again shows the UV excess towards the optical power-law model.
The power-law index $\alpha$ rises before the peak, and gradually decreases after the peak. This is consistent with the rise and fall of the blue continuum, which will be introduced in Section \ref{sec:specana}.

In MIR bands, five epochs of WISE W1 and W2 band photometry have been captured since the outburst, which reveal a prominent dust echo. A simple blackbody fitting indicates that the luminosity of the echo has risen to a peak level of log $L_{\rm bb,MIR}$ (erg s$^{-1}$) $\approx$ 43.3 in the second epoch, and only got weaken by 0.06 dex in the following three epochs; The temperature $T_{\rm bb,MIR}$ reaches $\approx$ 1,500 K in the first epoch and cools down to $\approx$ 900 K in the following epochs; The radius $R_{\rm bb,MIR}$ has expanded from $\approx4\times10^{16}$ cm (0.014 pc) to $\approx2\times10^{17}$ cm (0.066 pc). 

\subsection{X-ray Analysis}
The X-ray emission did not reach the detection threshold in the first 23 epochs (MJD 59768 to 60064). The total exposure time of all these epochs is 33.1 ks. We stack these epochs and detect a low average count rate of $(7.7\pm2.2)\times10^{-4}$ count s$^{-1}$. The first epoch that passes the threshold was on MJD = 60064, which was $\sim$250 days after the optical peak in the rest frame, and also contemporary to the plateau in the optical/UV bands. The average photon count rate for this epoch is $5.4^{+1.9}_{-1.5}\times10^{-3}$ count s$^{-1}$, which is $\sim$7.5 times higher than that of previous epochs. In order to improve the SNR, we bin the neighboring epochs and calculate the average rate for each bin. The binned light curve reveals a $\sim$60-day brightening phase following by a decline phase. The peak locates at MJD $\sim$ 60091, which was $\sim$280 rest-frame days after the optical peak, with the photon count rate of $(8.1\pm1.2)\times10^{-3}$ count s$^{-1}$.

For the early low state, we construct a stacked spectrum using the first 23 XRT epochs, and use \texttt{xspec} to analyze it. We first assume that the intrinsic spectrum is redshifted with $z=0.0735$, and passes through the Galactic absorption, with the hydrogen column density fixed at $N_{\rm H}=1.24\times10^{20}\ {\rm cm}^{-2}$ \citep{HI4PI2016}. Then, we fit the intrinsic spectrum into four models: a power-law profile (\texttt{tbabs*zashift*powerlaw}) and a blackbody model (\texttt{tbabs*zashift*bbody}), a multiple blackbody disk model (\texttt{tbabs*zashift*diskbb}) and a combination of a power-law profile and a multiple blackbody disk (\texttt{tbabs*zashift*(powerlaw+diskbb)}). All first three models yield good fit ($Cstat$/dof $<$ 1), while the last one cannot, since the total photon counts that build the spectrum is too low to constrain a large number of parameters. 
The spectrum can be well described by a power-law profile with $\Gamma=4.37^{+0.98}_{-0.80}$, a blackbody with $kT=100.8^{+28.7}_{-25.8}$ eV, or a multiple blackbody disk with temperature at the inner radius of $kT_{\rm in}=131.8^{+45.4}_{-37.7}$ eV. The Galactic-absorption-corrected rest-frame 0.3-10.0 keV (i.e., unabsorbed) luminosity, is $\sim$$10^{41.5}$ erg s$^{-1}$.

\begin{figure*}
    \centering
    \begin{minipage}{.3\linewidth}
        \subfigure[0.3-2.0 keV stacked image]{
            \includegraphics[width=3.5cm]{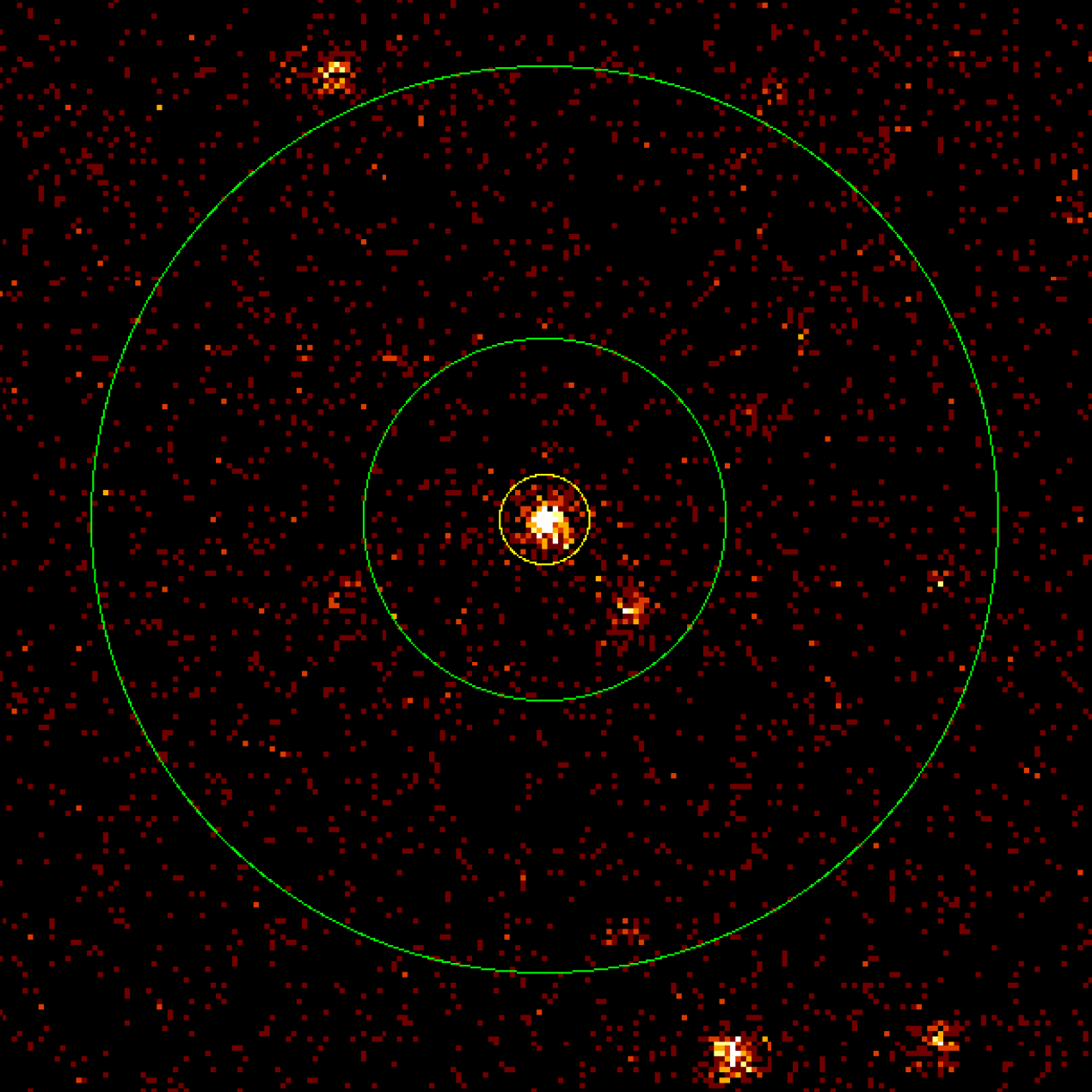}
        }
        \subfigure[2.0-10.0 keV stacked image]{
            \includegraphics[width=3.5cm]{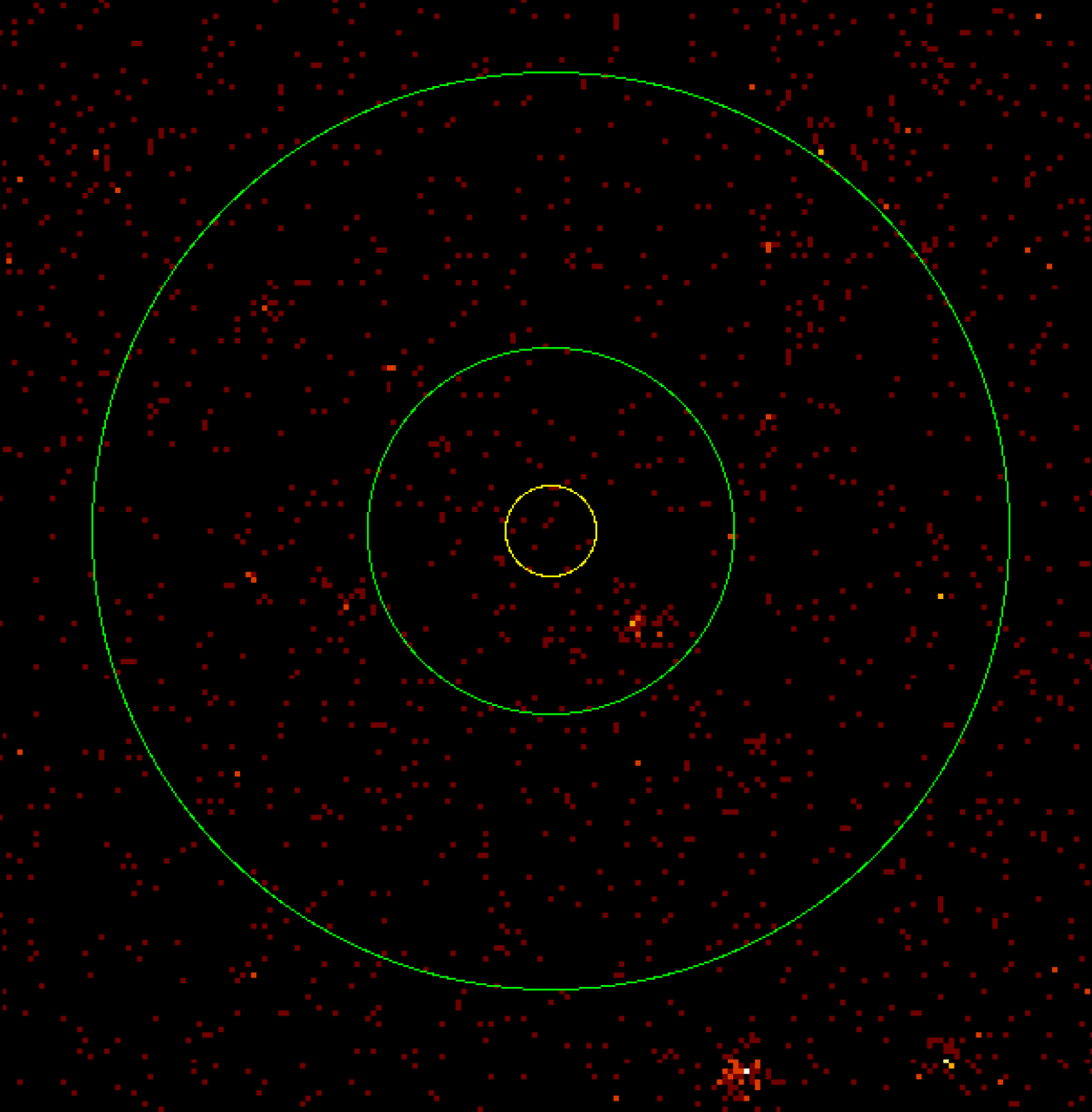}
        }
    \end{minipage}
    \begin{minipage}{.45\linewidth}
        \subfigure[Stacked spectra]{
            \includegraphics[width=8.8cm]{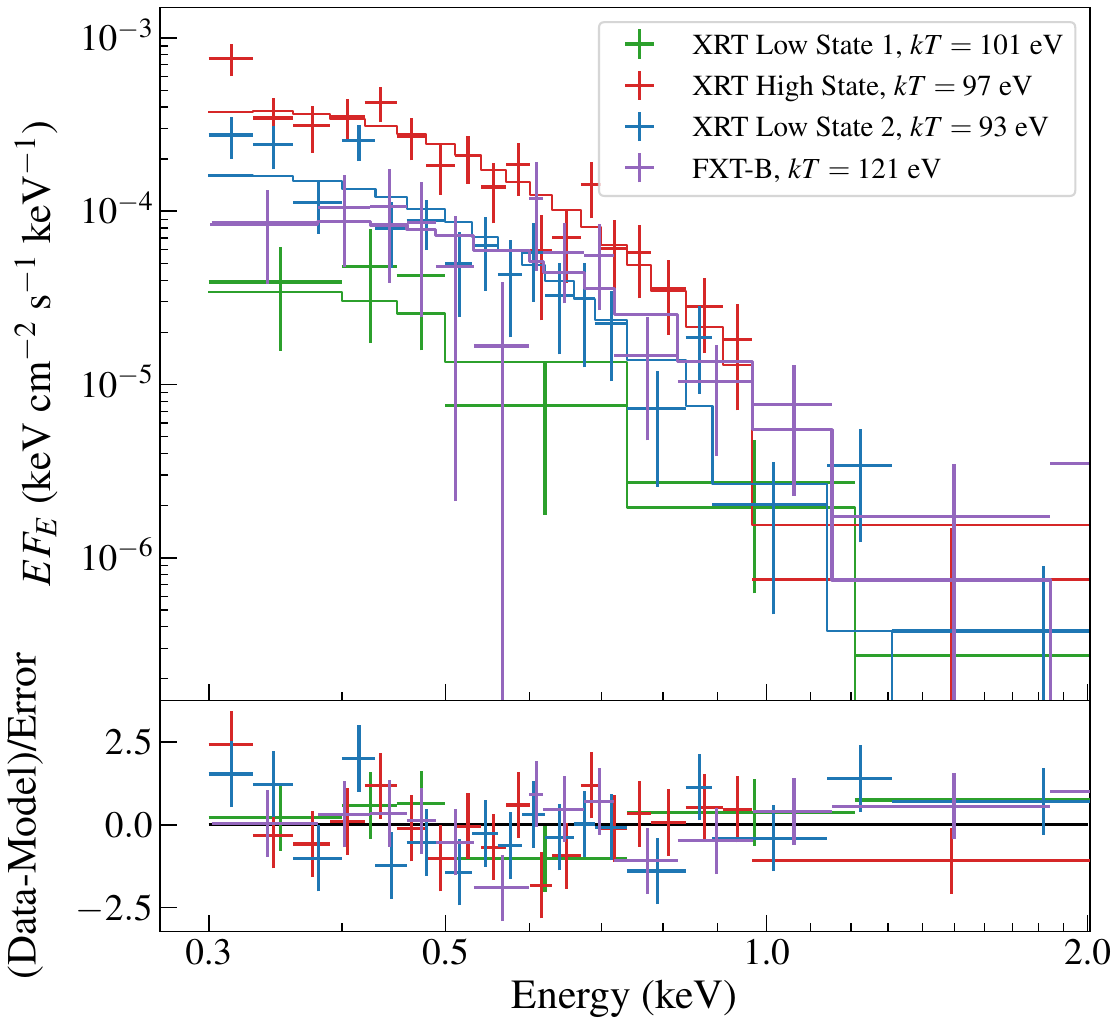}
    }
    \end{minipage}
    \caption{Swift XRT stacked images and spectra confirm an ultra-soft X-ray source. In panel (a) and (b), the stacked images of all epochs are shown, in which the source region is represented by the yellow circle with a radius of 20$^{\prime\prime}$, and the background region is represented by a green annulus with an inner radius of 80$^{\prime\prime}$ and an outer radius of 200$^{\prime\prime}$. In panel (c), the spectra and the best-fit blackbody models for three XRT phases and recent FXT-B observations are plotted. For clarity, only the soft X-ray (0.3-2.0 keV) wavelength range is plotted.}
    \label{fig:xrt}
\end{figure*}

In the post-peak epochs, after a quick drop in $\sim$80 days, the XRT photon count rate has been standing on a long plateau for $\sim$550 days. Therefore, we use this feature to define two 
phases: a ``high state'' (MJD 60071 to 60154) in which and ``low state 2'' (MJD 60176 to 60699). We construct a stacked spectrum for each phase, then we try to fit the intrinsic spectrum into the same four models. For the high state, the first three models yield good fit, while the last one cannot, since the \texttt{diskbb} component dominates and the \texttt{powerlaw} component cannot be constrained. The spectrum can be well described by a power-law profile with $\Gamma=4.87^{+0.28}_{-0.27}$, or a blackbody with $kT=97.1^{+4.7}_{-4.5}$ eV, or a multiple blackbody disk with $kT_{\rm in}=120.7^{+7.2}_{-6.7}$ eV. The unabsorbed luminosity is $\sim$$10^{42.6}$ erg s$^{-1}$. While for the low state 2, it can be well described by a power-law profile with $\Gamma=4.87^{+0.28}_{-0.27}$, a blackbody with $kT=92.8^{+6.5}_{-6.2}$ eV, or a multiple blackbody disk with $kT_{\rm in}=116.7^{+9.7}_{-9.0}$ eV. The unabsorbed luminosity is $\sim$$10^{42.2}$ erg s$^{-1}$. The EP FXT-B observations were carried out at the end of XRT low state 2. The stacked spectrum can be well described by a power-law profile with $\Gamma=3.67^{+0.40}_{-0.38}$, a blackbody with $kT=121.1^{+14.2}_{-12.3}$ eV, or a multiple blackbody disk with $kT_{\rm in}=159.9^{+24.9}_{-20.1}$ eV. The unabsorbed luminosity is $\sim$$10^{42.0}$ erg s$^{-1}$. To summarize, the X-ray spectrum of AT 2022fpx is ultra-soft, which is consistent with a bare disk with no corona or a blackbody with $kT\sim100$ eV. A summary of the fitting result is shown in Table \ref{tab:xrt}. In Figure \ref{fig:xrt}, panel (a) and (b) provide two stacked XRT images in soft and hard X-ray bands, which clearly shows that AT 2022fpx is an ultra-soft X-ray source, while panel (c) offers a direct view of the goodness of the blackbody fitting to the spectra.

\begin{table*}[htb!]
\footnotesize
\centering
\begin{threeparttable}\caption{Fitted parameters of the XRT \& FXT-B spectrum}\label{tab:xrt}
\doublerulesep 0.1pt \tabcolsep 8pt 
\begin{tabular*}{0.9\textwidth}{ccccccccc}
\toprule
  Phase & MJD & Exposure & Photon & Model & $\Gamma$ & $kT(T_{\rm in})$ & Unabsorbed log $L_{\rm X}$ & $Cstat$/dof \\\hline
    &59768&&&\texttt{powerlaw}&$4.37^{+0.98}_{-0.80}$&$-$&$41.68^{+0.16}_{-0.14}$&16.93/16\\
    XRT Low state 1 & $|$ & 33.1&21&\texttt{bbody}&$-$&$100.8^{+28.7}_{-25.8}$&$41.53^{+0.16}_{-0.15}$&18.55/16\\
    &60064&&&\texttt{diskbb}&$-$&$131.8^{+45.4}_{-37.7}$&$41.55^{+0.15}_{-0.16}$&18.07/16\\
    \hline
    &60071&&&\texttt{powerlaw}&$4.74^{+0.21}_{-0.21}$&$-$&$42.71^{+0.03}_{-0.04}$&75.83/57\\
    XRT High state&$|$&43.7&192&\texttt{bbody}&$-$&$97.1^{+4.7}_{-4.5}$&$42.55^{+0.04}_{-0.04}$&60.51/57\\
    &60154&&&\texttt{diskbb}&$-$&$120.7^{+7.2}_{-6.7}$&$42.58^{+0.04}_{-0.04}$&60.09/57\\
    \hline
    &60176&&&\texttt{powerlaw}&$4.87^{+0.28}_{-0.27}$&$-$&$42.32^{+0.05}_{-0.05}$&37.10/52\\
    XRT Low state 2 &$|$&73.8&101&\texttt{bbody}&$-$&$92.8^{+6.5}_{-6.2}$&$42.15^{+0.05}_{-0.04}$& 39.71/52\\
    &60699&&&\texttt{diskbb}&$-$&$116.7^{+9.7}_{-9.0}$&$42.18^{+0.05}_{-0.05}$&37.32/52\\
    \hline
    &60698&&&\texttt{powerlaw}&$3.67^{+0.40}_{-0.38}$&$-$&$42.15^{+0.09}_{-0.09}$&29.32/38\\
    FXT-B &$|$&6.9&39&\texttt{bbody}&$-$&$121.1^{+14.2}_{-12.3}$&$42.02^{+0.09}_{-0.09}$&26.68/38\\
    &60726&&&\texttt{diskbb}&$-$&$159.9^{+24.9}_{-20.1}$&$42.05^{+0.09}_{-0.09}$&26.81/38\\
\bottomrule
\end{tabular*}
\end{threeparttable}
\end{table*}

\subsection{Optical Spectral Analysis}\label{sec:specana}
\subsubsection{Emission line evolution} \label{sec:emiline}
\begin{figure*}[htb!]
    \centering
    \includegraphics[scale=0.4]{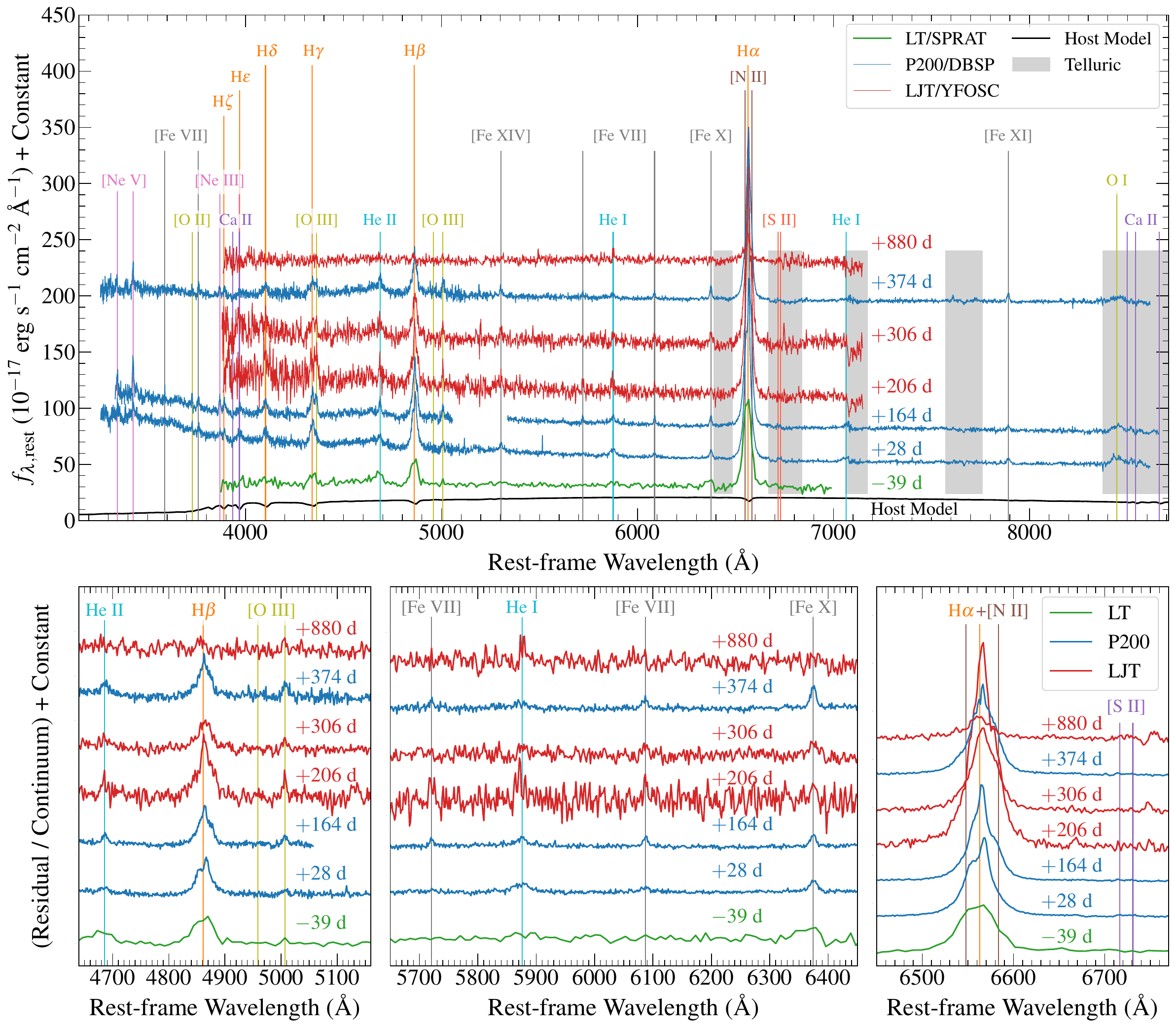}
    \caption{\textbf{Top panel:} All optical spectra. The black line represents the best modeled spectrum by \texttt{CIGALE} (See Section \ref{sec:sed} for details). Telluric absorption regions are marked in light gray. Each date stamp shows the rest-frame time difference between the optical peak (MJD $\sim$ 59795) and the date that the spectrum was taken. \textbf{Bottom panels:} Zoom-in views of He \textsc{ii}, H$\beta$ and [O \textsc{iii}] emission lines \textbf{(Left)}, He \textsc{ii} and iron coronal emission lines \textbf{(Middle)}, and blended H$\alpha$ and N \textsc{ii} emission lines \textbf{(Right)}.}
    \label{fig:spec}
\end{figure*}

As introduced in Section \ref{sec:spec}, 7 spectra have been taken for AT 2022fpx. As displayed in Figure \ref{fig:spec}, all spectra exhibit prominent Balmer, helium and iron coronal emission lines. For each spectrum, we fit the continuum by a power-law model or a third-degree polynomial, and subtract it to fit the emission lines. All three P200 spectra show a clear and fading excess in the wavelength of $<3900$ \AA\ that resembles a Balmer continuum.
In addition, a persistent excess in $\sim4400-4800$ \AA\ exists in all spectra, which can be explained by the Fe \textsc{ii} continuum. We try to fit both features with the method of \citet{Jin2012}, but the parameters cannot be well constrained, since the key features are blended with some emission lines, e.g., [Ne \textsc{v}] $\lambda\lambda3346,3426$, [O \textsc{ii}] $\lambda\lambda3726,3729$ and [Fe \textsc{vii}] $\lambda$3759 lines blend with the Balmer continuum; He \textsc{ii} $\lambda$4686 and [Fe \textsc{xiv}] $\lambda$5303 lines blend with the Fe \textsc{ii} continuum. Therefore, we only fit in the wavelength of $\geqslant3900$ \AA. In addition, we mask the telluric absorption regions (light gray regions in Figure \ref{fig:spec}) and skip fitting these severely affected lines: $[\,$S \textsc{ii}$\,]\,\lambda\lambda$6716, 6731; He \textsc{i}$\,\lambda7065$; O \textsc{i} $\lambda$8446; Ca \textsc{ii} triplet $\lambda\lambda$8498, 8542, 8662. The fitting results are shown in Table \ref{tab:spec1} and \ref{tab:spec2}. 

To summarize, the most prominent spectral features are: 
\begin{itemize}[noitemsep, topsep=0pt]
    \item A rise and fall of the blue continuum in $\sim$1 year.
    \item A series of strong Balmer emission lines (H$\alpha$, H$\beta$, H$\gamma$, H$\delta$) with luminosity of a few $\times$ 10$^{40-41}$ erg s$^{-1}$. Their FWHMs maintain at $\sim2000$ km s$^{-1}$;
    \item Helium emission lines (He \textsc{ii}$\,\lambda$4686 and He \textsc{i} $\lambda$5876) with luminosity of a few $\times$ 10$^{40}$ erg s$^{-1}$;
    \item Several iron coronal lines (e.g., [Fe \textsc{vii}]$\,\lambda$5721, 6087; [Fe \textsc{x}]$\,\lambda$6375; [Fe \textsc{xi}]$\,\lambda$7892; [Fe \textsc{xiv}]$\,\lambda$5303) with luminosity of $\sim$$10^{40}$ erg s$^{-1}$ and FWHMs of $(3-8)\times10^2$ km s$^{-1}$.
    \item The existence of Bowen fluorescence emission lines is ambiguous, as their wavelength ranges unfortunately overlap with those of other emission lines that should exist, e.g., N \textsc{iii} $\lambda\lambda$4100, 4640 are blended with the H$\delta$ ($\lambda4101$) and He \textsc{ii} $\lambda$4686 emission lines, respectively, while O III $\lambda$3760 overlaps with [Fe \textsc{vii}] $\lambda$3759. Hence, it is not certain whether AT 2022fpx is a Bowen fluorescent flare (BFF, e.g., \citealt{Trakhtenbrot2019,Makrygianni2023}).
    \item The recent YFOSC spectrum reveals much weaker emission lines. Particularly, the intensity of H$\alpha$ emission line has got weaker by a factor of $\sim$6 in the past 1.5 years. 
\end{itemize}

\begin{table*}[htb!]
\footnotesize
\centering
\begin{threeparttable}\caption{Fitted parameters of the key emission lines\label{tab:spec1}}
\doublerulesep 0.1pt \tabcolsep 7pt 
\begin{tabular*}{\textwidth}{cccccccccc}
\toprule
    &\multicolumn{4}{c}{TNS 2022-06-23$^\wedge$}&\multicolumn{4}{c}{P200 2022-09-03}\\\hline
    Emission&flux&$L$&offset&FWHM&flux&$L$&offset&FWHM\\
    Line&$10^{-17}\,$erg s$^{-1}\,$cm$^{-2}$&$10^{40}$ erg s$^{-1}$&km s$^{-1}$&km s$^{-1}$&$10^{-17}\,$erg s$^{-1}\,$cm$^{-2}$&$10^{40}$ erg s$^{-1}$&km s$^{-1}$&km s$^{-1}$\\\hline
    H$\delta$$^{\bowtie}$&$-$&$-$&$-$&$-$&272$\pm$24&3.8$\pm$0.3&45$\pm$79&2104$\pm$197\\
    H$\gamma_b$$^\dagger$&$-$&$-$&$-$&$-$&569$\pm$25&8.0$\pm$0.3&221$\pm$39&2425$\pm$116\\
    H$\gamma_n$$^\dagger$&$-$&$-$&$-$&$-$&39$\pm$12&0.5$\pm$0.2&331$\pm$38&358$\pm$100\\
    $[\,$O \textsc{iii}$\,]\,\lambda$4363$^\dagger$&$-$&$-$&$-$&$-$&$-$&$-$&$-$&$-$\\
    He \textsc{ii}$\,\lambda$4686*$^{\bowtie}$&529$\pm$61&7.5$\pm$0.9&-548$\pm$151&3167$\pm$385&281$\pm$18&4.0$\pm$0.3&-104$\pm$50&1884$\pm$133\\
    H$\beta_b$&947$\pm$54&13.4$\pm$0.8&-16$\pm$61&2533$\pm$152&1283$\pm$29&18.1$\pm$0.4&66$\pm$20&2122$\pm$49\\
    H$\beta_n$&$-$&$-$&$-$&$-$&102$\pm$17&1.4$\pm$0.2&369$\pm$21&363$\pm$55\\
    $[\,$O \textsc{iii}$\,]\,\lambda$5007&$-$&$-$&$-$&$-$&46$\pm$14&0.7$\pm$0.2&33$\pm$79&556$\pm$186\\
    $[\,$Fe \textsc{xiv}$\,]\,\lambda$5303*&$-$&$-$&$-$&$-$&188$\pm$18&2.7$\pm$0.2&18$\pm$76&1865$\pm$185\\
    $[\,$Fe \textsc{vii}$\,]\,\lambda$5721&$-$&$-$&$-$&$-$&$-$&$-$&$-$&$-$\\
    He \textsc{i}$\,\lambda$5876&$-$&$-$&$-$&$-$&184$\pm$19&2.6$\pm$0.3&-52$\pm$80&1664$\pm$192\\
    $[\,$Fe \textsc{vii}$\,]\,\lambda$6087&$-$&$-$&$-$&$-$&39$\pm$12&0.5$\pm$0.2&-18$\pm$100&672$\pm$236\\
    $[\,$Fe \textsc{x}$\,]\,\lambda$6375&$-$&$-$&$-$&$-$&110$\pm$13&1.5$\pm$0.2&23$\pm$40&716$\pm$95\\
    H$\alpha_b$&3743$\pm$115&52.8$\pm$1.6&-106$\pm$21&2082$\pm$59&4899$\pm$24&69.2$\pm$0.3&-45$\pm$4&1811$\pm$10\\
    H$\alpha_n$&$-$&$-$&$-$&$-$&416$\pm$14&5.9$\pm$0.2&253$\pm$3&330$\pm$10\\
    $[\,$Fe \textsc{xi}$\,]\,\lambda$7892&$-$&$-$&$-$&$-$&100$\pm$7&1.4$\pm$0.1&92$\pm$21&829$\pm$52\\\hline\hline
    &\multicolumn{4}{c}{P200 2023-01-27}&\multicolumn{4}{c}{LJT 2023-03-13}\\\hline
    H$\delta$$^{\bowtie}$&258$\pm$21&3.6$\pm$0.3&-29$\pm$67&1844$\pm$163&599$\pm$147&8.5$\pm$2.1&248$\pm$262&2448$\pm$674\\
    H$\gamma_b$$^\dagger$&473$\pm$29&6.7$\pm$0.4&51$\pm$49&1678$\pm$117&$-$&$-$&$-$&$-$\\
    H$\gamma_n$$^\dagger$&$-$&$-$&$-$&$-$&233$\pm$70&3.3$\pm$1.0&-465$\pm$106&791$\pm$253\\
    $[\,$O \textsc{iii}$\,]\,\lambda$4363$^\dagger$&176$\pm$24&2.5$\pm$0.3&-79$\pm$37&810$\pm$80&514$\pm$94&7.3$\pm$1.3&-367$\pm$115&1440$\pm$302\\
    He \textsc{ii} $\lambda$4686*$^{\bowtie}$&194$\pm$13&2.7$\pm$0.2&21$\pm$27&910$\pm$64&229$\pm$40&3.2$\pm$0.6&-93$\pm$65&787$\pm$156\\
    H$\beta_b$&1185$\pm$29&16.7$\pm$0.4&125$\pm$18&1821$\pm$49&1339$\pm$85&18.9$\pm$1.2&192$\pm$61&2488$\pm$185\\
    H$\beta_n$&82$\pm$18&1.2$\pm$0.3&161$\pm$21&286$\pm$57&213$\pm$59&3.0$\pm$0.8&117$\pm$37&508$\pm$110\\
    $[\,$O \textsc{iii}$\,]\,$$\lambda$5007&100$\pm$16&1.4$\pm$0.2&18$\pm$40&526$\pm$93&134$\pm$25&1.9$\pm$0.4&-13$\pm$36&399$\pm$85\\
    $[\,$Fe \textsc{xiv}$\,]\,$$\lambda$5303*&$-$&$-$&$-$&$-$&246$\pm$66&3.5$\pm$0.9&-393$\pm$235&2064$\pm$582\\
    $[\,$Fe \textsc{vii}$\,]\,$$\lambda$5721&45$\pm$12&0.6$\pm$0.2&5$\pm$47&358$\pm$112&79$\pm$25&1.1$\pm$0.3&-21$\pm$74&490$\pm$175\\
    He \textsc{i} $\lambda$5876&167$\pm$26&2.4$\pm$0.4&-47$\pm$109&1484$\pm$261&145$\pm$28&2.1$\pm$0.4&-249$\pm$59&624$\pm$139\\
    $[\,$Fe \textsc{vii}$\,]\,$$\lambda$6087&67$\pm$14&0.9$\pm$0.2&19$\pm$43&429$\pm$102&$-$&$-$&$-$&$-$\\
    $[\,$Fe \textsc{x}$\,]\,\lambda$6375&99$\pm$16&1.4$\pm$0.2&-4$\pm$39&502$\pm$92&$-$&$-$&$-$&$-$\\
    H$\alpha_b$&6113$\pm$35&86.3$\pm$0.5&102$\pm$4&1755$\pm$13&6232$\pm$69&88.0$\pm$1.0&161$\pm$7&1785$\pm$24\\
    H$\alpha_n$&671$\pm$23&9.5$\pm$0.3&109$\pm$3&331$\pm$9&604$\pm$56&8.5$\pm$0.8&112$\pm$9&422$\pm$28\\
    $[\,$Fe \textsc{xi}$\,]\,\lambda$7892&81$\pm$9&1.1$\pm$0.1&-16$\pm$30&652$\pm$68&$-$&$-$&$-$&$-$\\
\bottomrule
\end{tabular*}
\begin{tablenotes}
\item[*] : The emission lines that are blended with the possible Fe \textsc{ii} continuum.
\item[$^{\bowtie}$] : The emission lines that are blended with the possible Bowen fluorescent emission lines.
\item[$^\dagger$] : H$\gamma$ $\lambda$4341 and $[\,$O \textsc{iii}$\,]\,\lambda$4363 lines are blended with each other.
\item[$^\wedge$] : The SNR for this spectrum is relatively low, emission lines with flux-to-error ratio $<$3 are not listed.
\item Subscript $b$ \& $n$: $b$ and $n$ represent emission lines with FWHM $>$1000 km s$^{-1}$ and $<$1000 km s$^{-1}$, respectively.
\end{tablenotes}
\end{threeparttable}
\end{table*}

\begin{table*}[htb!]
\footnotesize
\centering
\begin{threeparttable}\caption{Fitted parameters of the key emission lines (Continued)\label{tab:spec2}}
\doublerulesep 0.1pt \tabcolsep 7pt 
\begin{tabular*}{\textwidth}{cccccccccc}
\toprule
    &\multicolumn{4}{c}{LJT 2023-05-28}&\multicolumn{4}{c}{P200 2023-09-09}\\\hline
    Emission&flux&$L$&offset&FWHM&flux&$L$&offset&FWHM\\
    Line&$10^{-17}\,$erg s$^{-1}\,$cm$^{-2}$&$10^{40}$ erg s$^{-1}$&km s$^{-1}$&km s$^{-1}$&$10^{-17}\,$erg s$^{-1}\,$cm$^{-2}$&$10^{40}$ erg s$^{-1}$&km s$^{-1}$&km s$^{-1}$\\\hline
    H$\delta$$^{\bowtie}$&295$\pm$66&4.2$\pm$0.9&283$\pm$142&1380$\pm$350&130$\pm$21&1.8$\pm$0.3&120$\pm$86&1200$\pm$216\\
    H$\gamma_b$$^\dagger$&348$\pm$84&4.9$\pm$1.2&156$\pm$195&1846$\pm$511&427$\pm$31&6.0$\pm$0.4&333$\pm$73&2695$\pm$186\\
    H$\gamma_n$$^\dagger$&$-$&$-$&$-$&$-$&$-$&$-$&$-$&$-$\\
    $[\,$O \textsc{iii}$\,]\,$$\lambda$4363$^\dagger$&$-$&$-$&$-$&$-$&15$\pm$5&0.2$\pm$0.1&77$\pm$18&109$\pm$35\\
    He \textsc{ii} $\lambda$4686*$^{\bowtie}$&448$\pm$65&6.3$\pm$0.9&44$\pm$144&2250$\pm$376&344$\pm$21&4.9$\pm$0.3&3$\pm$48&1873$\pm$124\\
    H$\beta_b$&1163$\pm$58&16.4$\pm$0.8&154$\pm$44&1962$\pm$105&859$\pm$27&12.1$\pm$0.4&108$\pm$24&1701$\pm$57\\
    H$\beta_n$&$-$&$-$&$-$&$-$&$-$&$-$&$-$&$-$\\
    $[\,$O \textsc{iii}$\,]\,\lambda$5007&121$\pm$27&1.7$\pm$0.4&-27$\pm$58&558$\pm$138&112$\pm$18&1.6$\pm$0.3&65$\pm$44&547$\pm$101\\
    $[\,$Fe \textsc{xiv}$\,]\,\lambda$5303*&145$\pm$29&2.0$\pm$0.4&15$\pm$79&847$\pm$187&111$\pm$14&1.6$\pm$0.2&-2$\pm$59&980$\pm$137\\
    $[\,$Fe \textsc{vii}$\,]\,\lambda$5721&$-$&$-$&$-$&$-$&37$\pm$10&0.5$\pm$0.1&4$\pm$49&390$\pm$117\\
    He \textsc{i} $\lambda$5876&100$\pm$27&1.4$\pm$0.4&83$\pm$92&704$\pm$218&69$\pm$19&1.0$\pm$0.3&-204$\pm$181&1377$\pm$431\\
    $[\,$Fe \textsc{vii}$\,]\,\lambda$6087&70$\pm$23&1.0$\pm$0.3&-12$\pm$77&484$\pm$182&51$\pm$12&0.7$\pm$0.2&-44$\pm$59&531$\pm$139\\
    $[\,$Fe \textsc{x}$\,]\,\lambda$6375&134$\pm$29&1.9$\pm$0.4&12$\pm$77&741$\pm$185&136$\pm$12&1.9$\pm$0.2&5$\pm$23&536$\pm$54\\
    H$\alpha_b$&6028$\pm$78&85.1$\pm$1.1&216$\pm$8&1807$\pm$25&4493$\pm$23&63.4$\pm$0.3&173$\pm$4&1659$\pm$10\\
    H$\alpha_n$&611$\pm$72&8.6$\pm$1.0&113$\pm$11&523$\pm$39&141$\pm$9&2.0$\pm$0.1&157$\pm$4&164$\pm$10\\
    $[\,$Fe \textsc{xi}$\,]\,\lambda$7892&$-$&$-$&$-$&$-$&113$\pm$8&1.6$\pm$0.1&0$\pm$15&501$\pm$31\\\hline\hline
    &\multicolumn{4}{c}{LJT 2025-03-06$^{\wedge}$}&\multicolumn{4}{c}{}\\\hline
    H$\delta$$^{\bowtie}$&$-$&$-$&$-$&$-$&&&&\\
    H$\gamma_b$$^\dagger$&$-$&$-$&$-$&$-$&&&&\\
    H$\gamma_n$$^\dagger$&$-$&$-$&$-$&$-$&&&&\\
    $[\,$O \textsc{iii}$\,]\,$$\lambda$4363$^\dagger$&$-$&$-$&$-$&$-$&&&&\\
    He \textsc{ii} $\lambda$4686*$^{\bowtie}$&$-$&$-$&$-$&$-$&&&&\\
    H$\beta_b$&79$\pm$19&1.12$\pm$0.27&-318$\pm$100&893$\pm$240&&&&\\
    H$\beta_n$&$-$&$-$&$-$&$-$&&&&\\
    $[\,$O \textsc{iii}$\,]\,\lambda$5007&44$\pm$12&0.62$\pm$0.17&-83$\pm$51&387$\pm$122&&&&\\
    $[\,$Fe \textsc{xiv}$\,]\,\lambda$5303*&62$\pm$18&0.87$\pm$0.26&413$\pm$118&854$\pm$279&&&&\\
    $[\,$Fe \textsc{vii}$\,]\,\lambda$5721&$-$&$-$&$-$&$-$&&&&\\
    He \textsc{i} $\lambda$5876&86$\pm$14&1.21$\pm$0.2&134$\pm$43&546$\pm$102&&&&\\
    $[\,$Fe \textsc{vii}$\,]\,\lambda$6087&$-$&$-$&$-$&$-$&&&&\\
    $[\,$Fe \textsc{x}$\,]\,\lambda$6375&$-$&$-$&$-$&$-$&&&&\\
    H$\alpha_b$&747$\pm$26&10.55$\pm$0.37&-71$\pm$25&1541$\pm$60&&&&\\
    H$\alpha_n$&$-$&$-$&$-$&$-$&&&&\\
    $[\,$Fe \textsc{xi}$\,]\,\lambda$7892&$-$&$-$&$-$&$-$&&&&\\
\bottomrule
\end{tabular*}
\begin{tablenotes}
\item[*] : The emission lines that are blended with the possible Fe \textsc{ii} continuum.
\item[$^{\bowtie}$] : The emission lines that are blended with the possible Bowen fluorescent emission lines.
\item[$^\dagger$] : H$\gamma$ $\lambda$4341 and $[\,$O \textsc{iii}$\,]\,\lambda$4363 lines are blended with each other.
\item[$^\wedge$] : The SNR for this spectrum is relatively low, emission lines with flux-to-error ratio $<$3 are not listed.
\item Subscript $b$ \& $n$: $b$ and $n$ represent emission lines with FWHM $>$1000 km s$^{-1}$ and $<$1000 km s$^{-1}$, respectively.
\end{tablenotes}
\end{threeparttable}
\end{table*}

\section{Discussion: A SN, an AGN or a TDE?}
\label{sec:discuss}
In this section, we 
judge the possibilities for the three common origins of nuclear transients: a SN, a variable/turn-on AGN and a TDE, based on all observational facts.

\subsection{Transients for comparison}\label{sec:trans4comp} 
To confirm the type of AT 2022fpx, we compare its near-peak and late-time spectra with these transients that share some features with AT 2022fpx: 
\begin{itemize}[noitemsep, topsep=0pt]
\item AT 2019qiz, a typical TDE; 
\item AT 2019brs and AT 2019fdr, two outbursts in NLSy1 galaxies; 
\item AT 2018dyk, an outburst in a LINER galaxy; 
\item SDSS1335+0728, a turn-on AGN; 
\item SDSS J1115+0544, a turn-on AGN/TDE candidate; 
\item SDSS J0952+2143, an extreme coronal line emitter (ECLE); 
\item SN2008iy, a long-rise SN IIn; 
\item SN2010jl, a long-decline, X-ray- and IR-bright nuclear SN IIn. 
\end{itemize}
For their detailed information, please refer to Appendix \ref{sec:detinfo}. 

\begin{figure*}[htb!]
    \centering
    \includegraphics[scale=0.25]{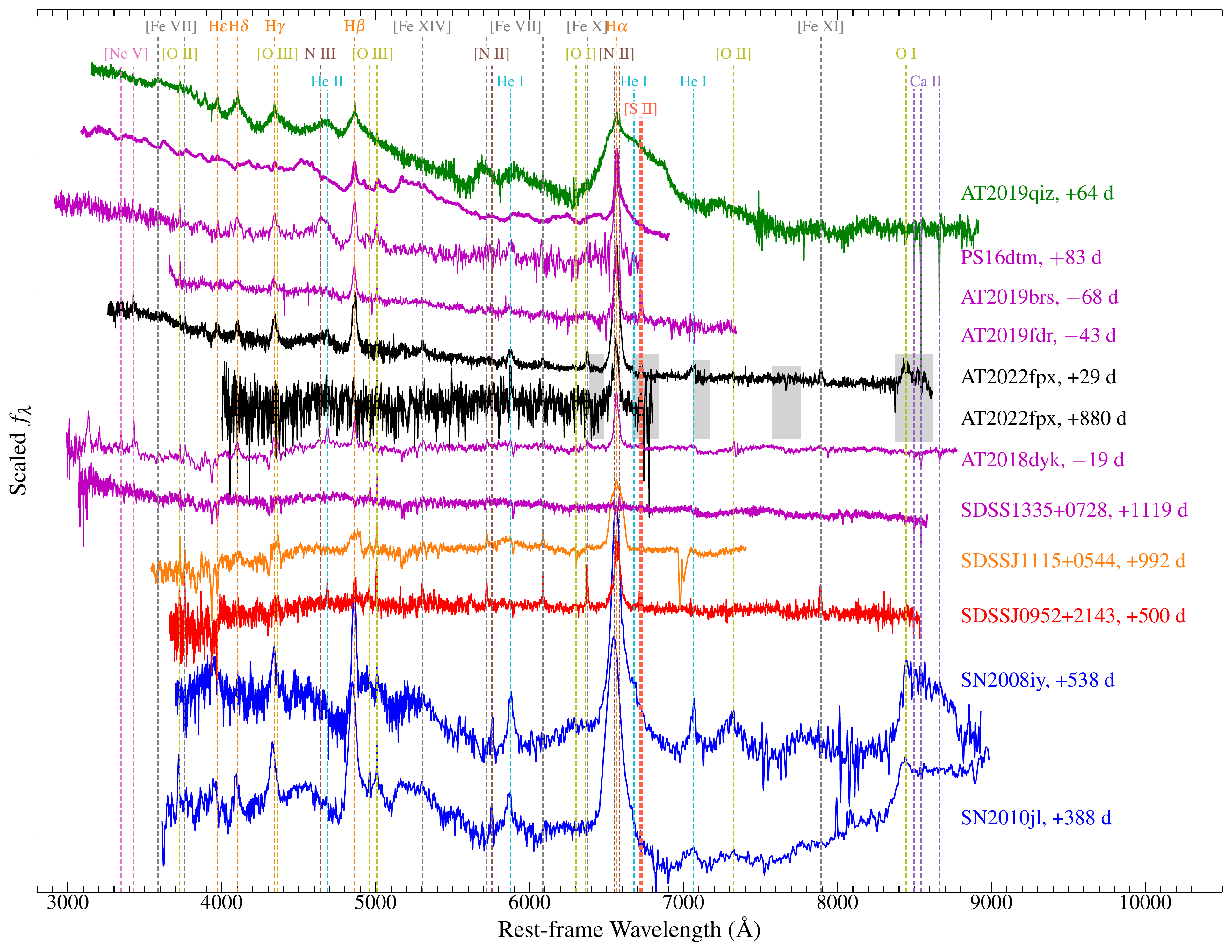}
    \caption{The spectral comparison between AT 2022fpx and a set of transients that share similarities: A typical TDE AT 2019qiz; Two outbursts in NLSy1 galaxies, AT 2019brs and AT 2019fdr; An outburst in a LINER galaxy, AT 2018dyk; A turn-on AGN SDSS1335+0728; A turn-on AGN/TDE candidate SDSS J1115+0544; An ECLE SDSS J0952+2143; A long-rise SN IIn SN2008iy; A long-decline, X-ray- and IR-bright nuclear SN IIn SN2010jl. Telluric regions are marked in gray.}
    \label{fig:comp}
\end{figure*}

We collect the spectra of these sources except for iPTF 16bco from WISeREP \citep{WISeREP}, \citet{Sanchez2024} and our follow-up observational data, then plot them with the near-peak and late-time spectra of AT 2022fpx in Figure \ref{fig:comp}.

\subsection{SN: Unlikely}\label{sec:onSN}

First, we try to explain the facts in a SN scenario. 

The absolute magnitude of $M_r\sim-20$ does not reach the threshold of $M_r<-21$ for a superluminous supernova (SLSN), unless a strong host extinction is taken into account. Since all the optical spectra show prominent Balmer and helium emission lines and absence of absorption features, the source is most likely to be a SN IIn. The FWHM of $\sim$2000 km s$^{-1}$ is well consistent with the typical SN IIn. Meanwhile, its red color is well acceptable for a SN, and the lack of color evolution is shown in a great number of SNe IIn \citep{Nyholm2020}.

However, AT 2022fpx shows some features that are atypical for a SN IIn. First, the rise phase of $\sim120$ days is the second longest for a SN IIn, only shorter than SN2008iy. Second, the late-time soft X-ray emission brightening did not change the spectral profile, reached a peak unabsorbed luminosity of log $L_{\rm X}\approx10^{42.6}$ erg s$^{-1}$, then maintained at a plateau of log $L_{\rm X}\approx10^{42.2}$ erg s$^{-1}$. This soft X-ray behavior is unprecedented for any SN IIn. Third, the persistent existence of extreme iron coronal lines ([Fe \textsc{vii}] to [Fe \textsc{xiv}]) of $\sim10^{40}$ erg s$^{-1}$ requires a central engine that constantly produces X-ray photons of $E\gtrsim100$ eV. Although coronal emission lines can be produced by some SNe IIn, e.g., SN2006jd \citep{Stritzinger2012} and SN2010jl \citep{Fransson2014}, they emerge several hundred days post-peak, and the luminosity is $\lesssim10^{38}$ erg s$^{-1}$ \citep{Wang2023}. Fourth, the luminosity of the dust echo is $L_{\rm MIR}\approx2\times10^{43}$ erg s$^{-1}$, which exceeds those of most SNe IIn by one to two order(s) of magnitude ($10^{8-9}\,L_\odot$, \citealt{Fox2013,Wang2018,Jiang2019,Szalai2019}), and comparable with two brightest echoes in SN2003ma \citep{Rest2011} and SDWFS-MT-1 \citep{Kozlowski2010}. Fifth, the nebular lines are not shown in the late-time spectrum. As shown in Figure \ref{fig:comp}, despite some key lines are severely contaminated by the telluric absorption, e.g., O \textsc{i} $\lambda$8446; Ca \textsc{ii} triplet $\lambda\lambda$8498, 8542, 8662, the lack of nebular
[O \textsc{ii}]$\,\lambda\lambda$7320, 7331 still reflects the problem. Although the great diversity among the SN IIn family makes a clear exclusion difficult, we conclude that AT 2022fpx is unlikely to be powered by a SN IIn.

\subsection{Turn-on AGN vs. TDE}\label{sec:turnonAGNorTDE}
Now that we have excluded the SN scenario, we turn to a more difficult problem, that is to judge if AT 2022fpx is more likely to be a turn-on AGN, or a long-lived TDE. We investigate this problem in four aspects.

\subsubsection{Pre-existing AGN \& Emission line profile}
The unfortunate lack of pre-outburst spectrum mounts the difficulty of judging the existence of a pre-existing AGN. In Section \ref{sec:wise} and \ref{fig:preflare}, we conclude that a weak AGN probably existed before the outburst by the weak MIR variability. We further look into the comparison with other nuclear transients, and find additional supportive evidence for this conclusion. 
First, the Fe \textsc{ii} continuum persistently exists in all optical spectra, which is usually presented in AGNs. Second, as displayed in Figure \ref{fig:comp}, the line profile of AT 2022fpx exhibit pure emission of Balmer, He \textsc{i}, He \textsc{ii}, [O \textsc{ii}], [O \textsc{iii}], Fe \textsc{ii} continuum and extreme iron coronal emission lines, which exactly combines those of outbursts in NLSy1 galaxies, AT 2019brs and AT 2019fdr, with those of ECLEs, such as SDSS J0952+2143, but different from those of the turn-on AGN SDSS1335+0728. The FWHM of H$\alpha$ and H$\beta$ maintains at $\sim2000$ km s$^{-1}$, which largely differs from the typical TDE AT 2019qiz, but again resembles those of outbursts in NLSy1 galaxies and ECLEs, whose FWHMs maintain at $\sim$$1000-2000$ km s$^{-1}$. Hence, AT 2022fpx is most likely to be a transient occurring around a weakly active BH.



\subsubsection{Optical-UV luminosity evolution}\label{sec:optuvevo}
After confirming the pre-existing weak nuclear activity, we continue to explore the origin of the recent outburst.
The inset of the Figure \ref{fig:lc} displays a precursor before the main rise stage. While precursors have been reported to exist in the TDE light curves, this precursor is spiky in the light curves and disappeared before the main rise stage, which is different from those in published literature, e.g., the long-lasting precursor in AT 2023lli \citep{Huang2024}, the bump during the rise stage in AT 2018gn \citep{Wang2024}, AT 2019azh \citep{Faris2024} and AT 2020wey \citep{Charalampopoulos2023}. We note that a TDE-in-AGN candidate reported in a recent preprint, AT 2024kmq, has a highly similar red precursor \citep{Ho2025}.

The long rise stage of AT 2022fpx resembles those of AT 2019brs and AT 2019fdr \citep{Frederick2021}. In Figure 10 of \citet{Frederick2021}, the authors set 9 criteria to judge if a TDE or an AGN flare is favored. Briefly speaking, a TDE is favored when $M_{\rm BH}<10^8\,M_\odot$, H$\beta$ FWHM $>2200$~km~s$^{-1}$, [O \textsc{iii}]/H$\beta$ flux ratio $>3$ (above two are based on opposing the NLSy1 selection criteria of \citealt{Rakshit2017}), no Fe \textsc{ii} continuum presents, $g-r<0$, UV emission is persistent, X-ray is weak or $\Gamma\gtrsim5$, host MIR color W1 $-$ W2 $<0.7$, and no rebrightening flare. Using this criteria, the authors voted AT 2019brs as an AGN flare and AT 2019fdr as a TDE. Regarding that TDEs also commonly have low [O \textsc{iii}]/H$\beta$ ratio, and rebrightening flares have been found in some TDEs (or candidates), we discard criteria of the [O \textsc{iii}]/H$\beta$ flux ratio and rebrightening flare, and employ the remaining criteria on AT 2022fpx. A TDE is marginally favored (4 vs. 3). 

Despite this, the rise timescale is much longer than those of the typical optical TDEs: In Section \ref{sec:flare}, we derive the Gaussian rise timescale for the $g$ band as $\sigma=41.5$ d, which is $\sim4$ times higher than the predicted $\sigma$ for $M_{\rm BH}\sim 10^6\,M_\odot$ \citep{Huang2023}. 
We also derive the $t_{\rm 1/2,rise}$, $t_{\rm 1/2,decline}$ and $t_{\rm 1/2}$ timescales following the method that \citet{Yao2023} applied to the ZTF TDE sample, and find that all of them fall on the longest side in the ZTF TDE sample, and in particular, the $t_{\rm 1/2}$ is $>$3 times to those TDEs with $M_{\rm BH}\sim 10^6\,M_\odot$.
We attempt to solve this discrepancy as follows.
First, we regard that both $\sigma$ and $t_{\rm 1/2}$ are related to the fallback timescale $t_{\rm fb}$, and $t_{\rm fb}\propto M_{\rm BH}^{1/2}M_\star^{-1} R_\star^{3/2}\beta^{-3}$ \citep{Lodato2011}, where $M_\star$ and $R_\star$ are the mass and radius of the disrupted star, and the impact parameter $\beta$ is defined as the ratio of the tidal radius and the pericenter, $\beta=R_{\rm t}/R_{\rm p}$. If the disrupted star is a main-sequence (MS) star, which roughly obeys $R_\star\propto M_\star^{0.8}$ \citep{Stone2016}, then $t_{\rm fb}\propto M_{\rm BH}^{1/2}M_\star^{1/5}\beta^{-3}$, relying very weakly on the mass of the disrupted star. However, if the star is a giant star with much lower density, e.g., $M_\star\sim1.5\,M_\odot$ and $R_\star\sim10\,R_\odot$, then a $\sim$20 times longer $t_{\rm fb}$ can be achieved \citep{Macleod2012,Law-Smith2017}. In addition, $t_{\rm fb}$ is highly influenced by the value of $\beta$. Under a MS star assumption, $\beta$ should be lower than average by a factor of $\sim1.4-1.6$ to solely explain the factor of 3-4. While under a giant star assumption, $\beta$ should be higher than average by a factor of $\sim1.7-1.9$. \citet{Nicholl2022} fitted 32 optical TDEs by \texttt{MOSFiT} \citep{Guillochon2018} and found the average $\left<\beta\right>\gtrsim1.1$ for TDE-H+He. Therefore, MS star assumption requires $\beta\sim0.7-0.8$, which is a very shallow partial TDE. However, it is strongly against the long-lasting and energetic appearance of this transient. While giant star requires $\beta\sim1.9-2.1$, which appears to be more self-consistent.

Another key discrepancy is the red optical color of $g-r>0$, which can be either intrinsic or caused by a heavy dust reddening. We will investigate this problem in Section \ref{sec:red}.

To summarize, based on the optical-UV luminosity evolution, a TDE is slightly favored, but a giant star is probably required to produce such a long rise timescale.


\subsubsection{Ultra-soft X-ray emission}
AT 2022fpx displays a stable and ultra-soft X-ray profile in both high and low state, which can be described by a $kT\sim100$ eV blackbody or a power-law with photon index $\Gamma\sim4-5$. The panel (a) and (b) Figure \ref{fig:xrt} clearly demonstrate its softness, as it is outstandingly bright within 0.3-2.0 keV but much fainter than the nearby X-ray sources within 2.0-10.0 keV. The X-ray source is unlikely to be a NLSy1 accretion disk with $\Gamma\sim3$ \citep{Boller1996, Rakshit2017}, and even less likely to be a typical AGN accretion disk with $\Gamma\sim1.75$ \citep{Ricci2017}, but most likely to be
a nascent TDE accretion disk. 

The X-ray peak lags $\sim$280 rest-frame days behind the optical-UV peak.
This time lag is often ascribed to a TDE, e.g., \citet{Gezari2017b, Pasham2017, Shu2020, Liu2022, Wang2022b, Huang2023}, under two competing scenarios. The first scenario is that 
the UV/optical emission starts producing once the self-intersection of the tidal debris begins, while X-ray emission can only start once the accretion disk forms. The second one is that the X-ray photons are initially reprocessed into UV/optical emission due to the obscuration of some material, and can only pass through once the material get optically thin. 

For the former case, the time lag corresponds to the timescale of circularization. For a solar mass disrupted star, $t_{\rm cir}=340.3\,(M_{\rm BH}/10^6\,M_\odot)^{-7/6}\beta^{-3}$ d \citep{Gezari2017b}. Considering $M_{\rm BH}\sim10^{6.24}\ M_\odot$ and $\beta\sim1$, the timescale is well consistent with the observed $\sim$280-day time lag. 
However, if $\beta\sim2$, the timescale will be much shorter than the observed value.
More importantly, extreme coronal emission lines that persistently exhibit in all spectra, including those taken during the optical-UV rise and X-ray low stage, suggest that a large number of X-ray photons had been presented before the rebrightening phase, therefore should support the reprocessing scenario instead the self-intersection one. The prominent dust echo and extreme neon and iron coronal lines give additional evidence towards a dusty and gas-rich circumnuclear environment. 

The X-ray luminosity has been maintaining at $\sim$$10^{42}$ erg s$^{-1}$ for $\sim$550 d. It contradicts the typical $t^{-5/3}$ decline trend for TDEs, but similar plateaus have been found in some optical-selected TDEs \citep{Guolo2024}. Regarding that its ultra-soft spectrum is unusual for AGNs, follow-up X-ray observations are still needed to judge its nature.

\subsubsection{Optical-red: Intrinsic, contaminated by emission lines, or heavily dust-reddened?} \label{sec:red}


The constantly red optical color of $g-r\sim0.4$ contradicts the current TDE selection criterion of $g-r<0$. The criterion is based on the past experience that the SED of optical TDEs can be well fitted by a blackbody of $T_{\rm bb}\sim(1-5)\times10^4$ K. However, AT 2022fpx contradicts this deduction, as it shows a red optical color.

We first consider the case that the optical red color is intrinsic. When fitted with only optical bands, the blackbody temperature is only $\sim$7500 K. If UV bands are included, the temperature rises to $\sim$14000 K, but the reduced $\chi^2$ is much higher than 1, indicating that the optical-UV SED is not conform to a blackbody model. Although the power-law model also shows discrepancy, the best-fit model fits the optical-UV SED well. The power-law index of $\sim$1.58 is similar to that of the turn-on AGN iPTF 16bco ($f_{\lambda}\propto\lambda^{-1.45}$) and common AGNs ($f_{\lambda}\propto\lambda^{-1.5}$, Section 1.1 of \citealt{Netzer2013}). 

We make a second guess. Regarding that the spectra exhibit prominent emission lines, and the most prominent line, H$\alpha$, lies in the wavelength range of the $r$ band, it is possible that their contamination will significantly enhance the $r$-band luminosity and result in a ``false'' red color. The recent fading of H$\alpha$ emission line and the bluer color may support this assumption. 
In an extreme case, all H$\alpha$ emission only appears post-peak, yielding a reddest color. We evaluate the contribution of the continuum and emission lines by fitting three high-SNR P200 spectra. The continuum is described by the sum of the host galaxy and a third-degree polynomial, $f_{\lambda}=C_{\rm host}f_{\lambda,{\rm host}}+\sum_{i=0}^3 C_i\lambda^i$, where $C$ represents constant parameter, and the residual is regarded as the emission line profile. Using the ZTF-$g$ and $r$ throughput curves to calculate the $g-r$ color of the third-degree polynomial with or without the companion of the emission line profile. We find that although the emission lines can redden the color by $\sim$0.2 mag at most, the continuum components for all spectra have intrinsically red color of $g-r>0$, therefore the contamination of emission lines is not the main reason behind the red color.

\begin{figure*}[]
    \centering
    \subfigure[Blackbody fitting results]{
    	\includegraphics[width=8cm]{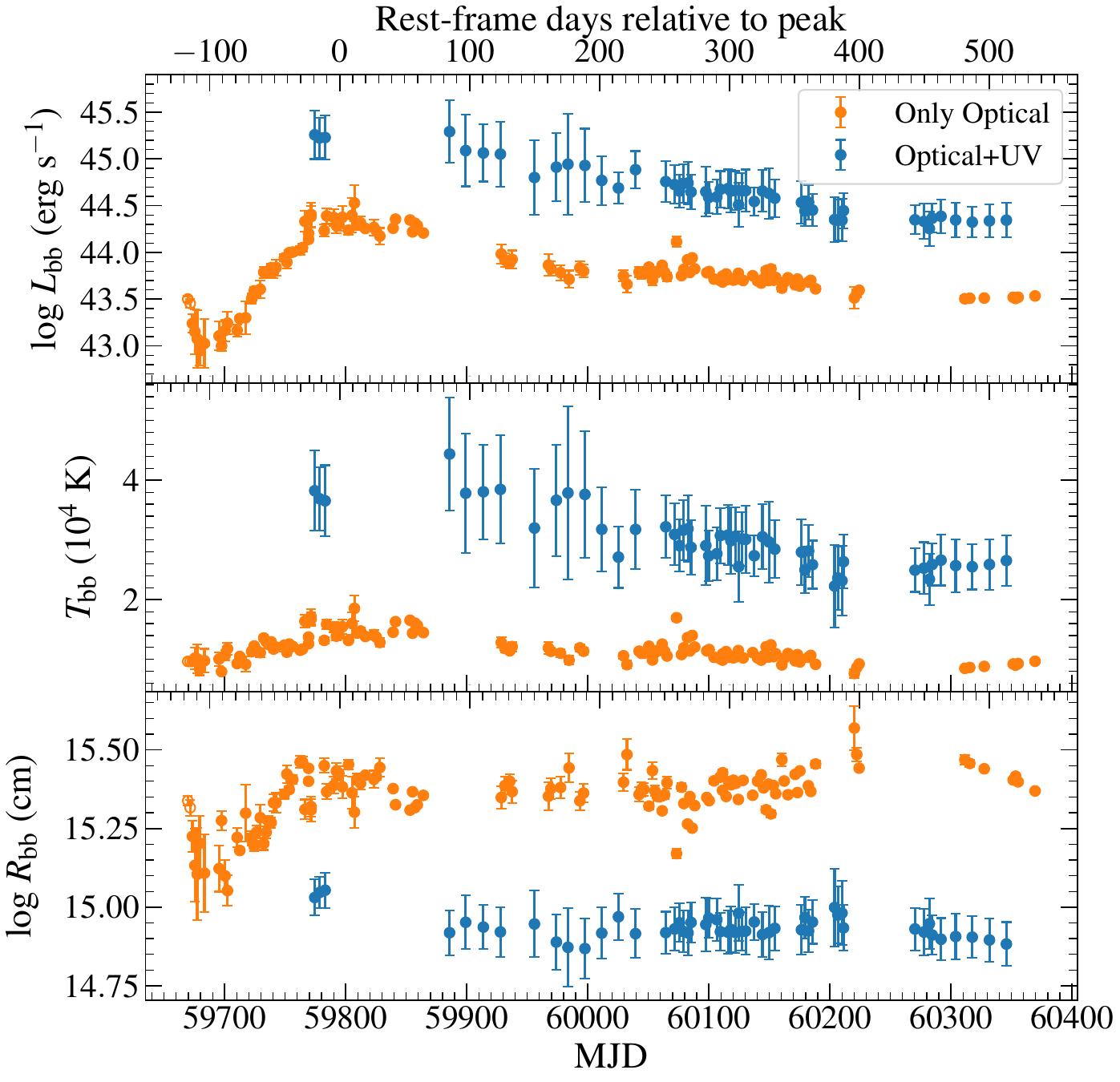}
    }
    \centering
    \subfigure[Power-law fitting results]{
    	\includegraphics[width=8cm]{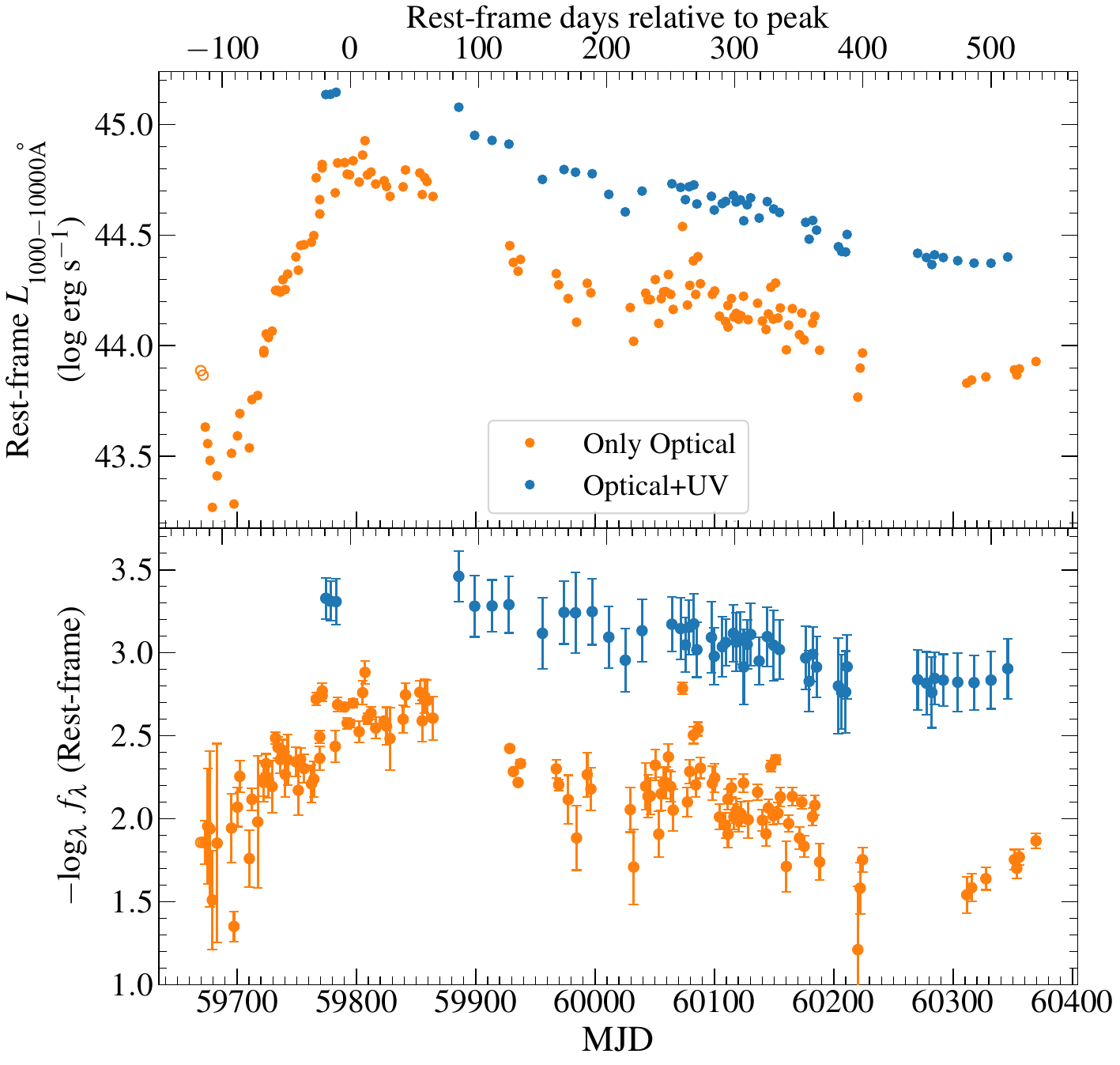}
    }
    \caption{The fitting results for UV and optical photometric data after a host extinction correction of $A_V\sim1.14$ mag. \textbf{Left:} The blackbody luminosity $L_{\rm bb}$, temperature $T_{\rm bb}$ and radius $R_{\rm bb}$. Since the first two epochs only have one band photometry, their blackbody temperatures are fixed as the best-fit temperature of the third epoch, so as the power-law indexes. \textbf{Right:} The rest-frame 1000$-$10000 \AA\ luminosity and the power-law index $\alpha$ ($f_{\lambda}\propto\lambda^{-\alpha}$, rest-frame).}
    \label{fig:bb+pox}
\end{figure*}

Third, we check if a strong host galaxy extinction can redden a TDE-like SED into a similar shallow SED.
The spectral fitting reveals that H$\alpha$/H$\beta$ line ratio rises from $\sim$4 (early stage) to $\sim$5 (late-time) (See Table \ref{tab:spec1}). To estimate the dust attenuation factor, we first evaluate the intrinsic H$\alpha$/H$\beta$ line ratio by the \texttt{PyNeb} package \citep{Luridiana2015}. Assuming a case B recombination scenario and typical values for broad line region of $T_e=10^4$~K, $n_e=10^9$~cm$^{-3}$, the intrinsic H$\alpha$/H$\beta$ line ratio is 2.62. We test on a grid of log $T_e$ = [4.0, 4.4] and log $n_e$ = [7, 10] and find intrinsic values lie in a small range of [2.61, 2.77]. Then we adopt the extinction law from Fitzpatrick (1999) with $R_V=3.1$ which is suitable for emission lines \citep{LinZS2024}.
When H$\alpha$/H$\beta$ = 4, $A_V\sim1.14$ mag.
When H$\alpha$/H$\beta$ = 5, $A_V\sim1.73$ mag.
To investigate if the host extinction is enough to cause the optical red color, we apply the more conservative factor 
$A_V\sim1.14$ mag to the SED and perform another blackbody fitting. After correction, the optical color and the optical blackbody temperature appear more reasonable in a TDE frame ($g-r<0$ and $T_{\rm bb,opt}\sim13000$ K). However, the discrepancy becomes larger as $T_{\rm bb,opt+UV}\sim40000$ K, although both temperatures lie in the acceptable range of TDE blackbody temperature. We also perform a power-law fitting on the corrected SED, and find the power-law index for the near-peak SED rises from $\sim$2.5 to $\sim$3.3 if UV bands are included. In Figure \ref{fig:bb+pox} we show the fitting results for the full optical-UV light curves.

Regarding that the color turns into blue ($g-r<0$) after the correction, and no previous work has examined that if TDEs show the same discrepancy of blackbody temperature and power-law index between the inclusion or exclusion of UV bands, the possibility of a TDE cannot be ruled out. We further look into this problem by comparing the SED of AT 2022fpx with those of the typical TDEs that are best-sampled in optical and UV bands. The details are introduced below.

\begin{figure*}[t]
\centering
\subfigure[AT 2022fpx]{
	\includegraphics[width=5.5cm]{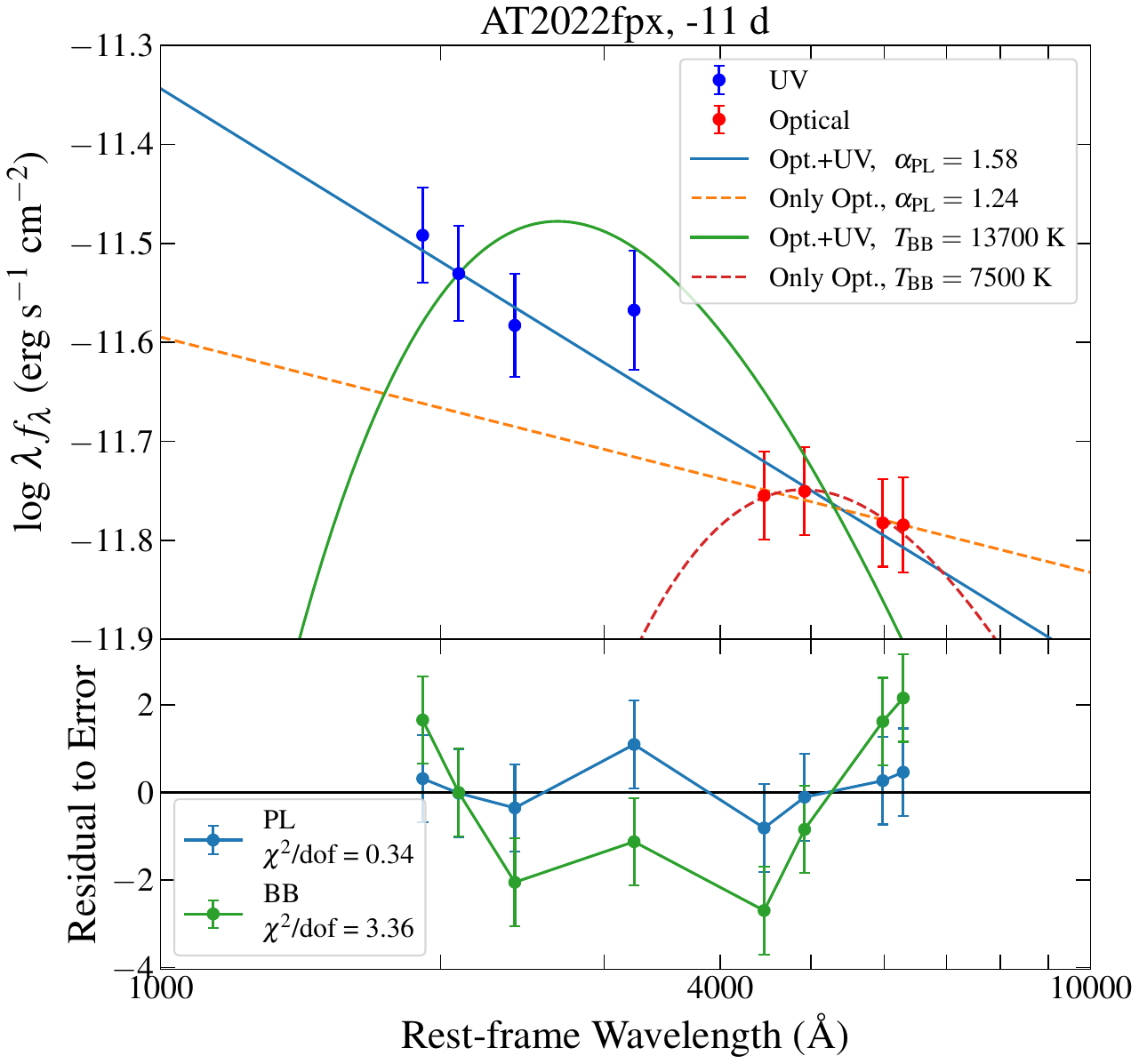}
}
\subfigure[AT 2022fpx (Host extinction corrected)]{
	\includegraphics[width=5.5cm]{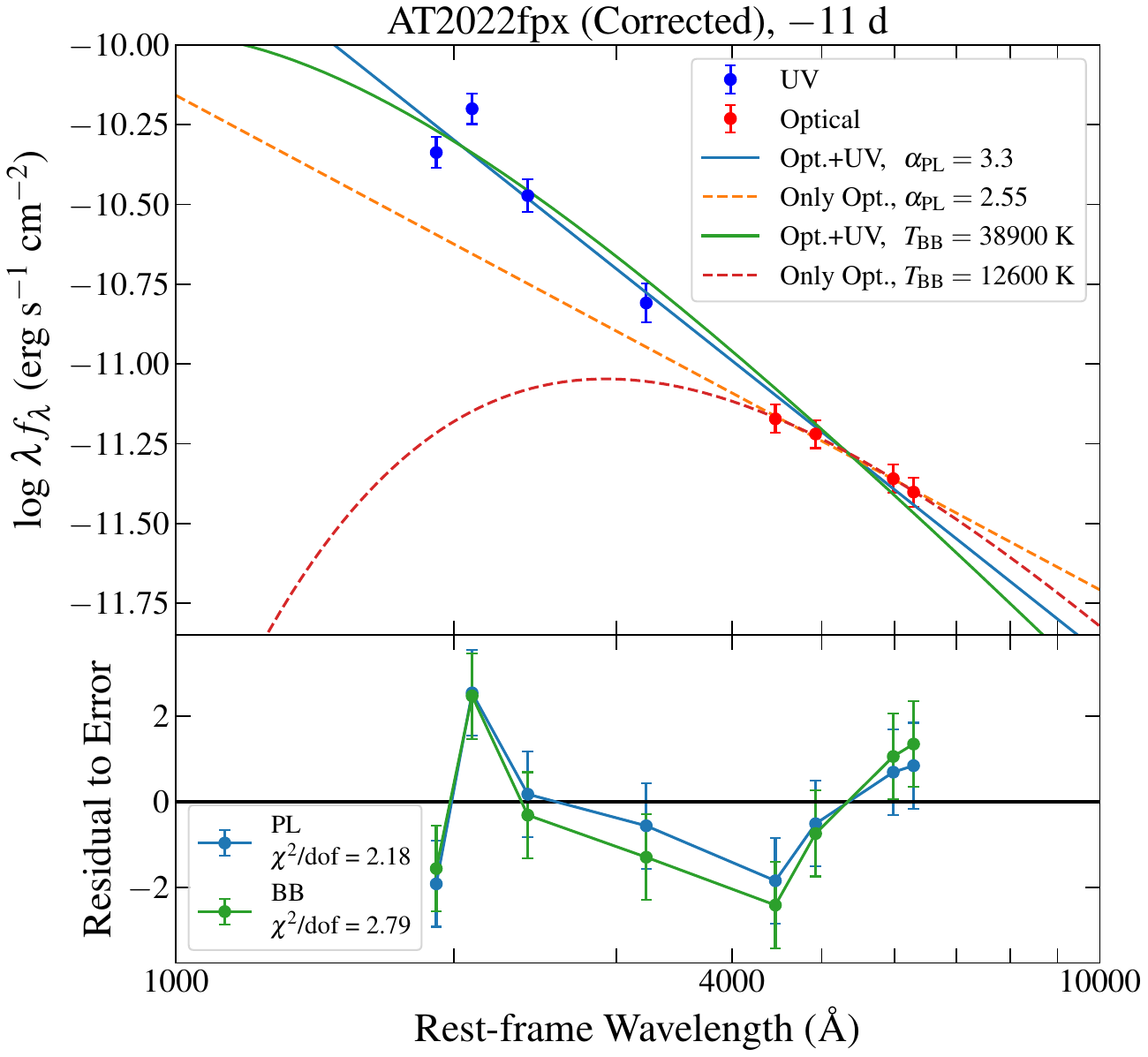}
}
\subfigure[AT 2018dyb]{
	\includegraphics[width=5.5cm]{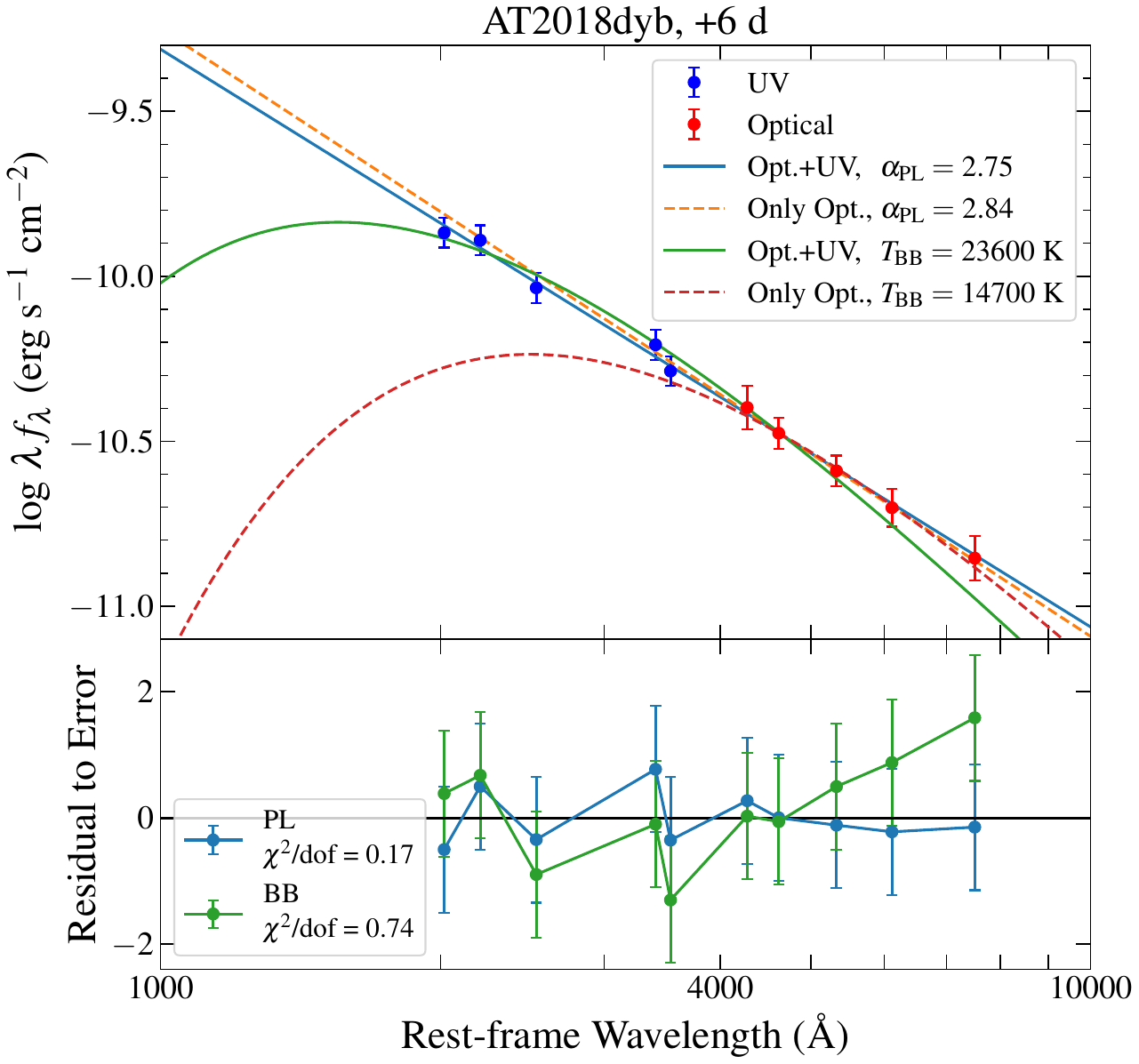}
}\\
\hspace{0.5pt}
\subfigure[AT 2019azh]{
	\includegraphics[width=5.5cm]{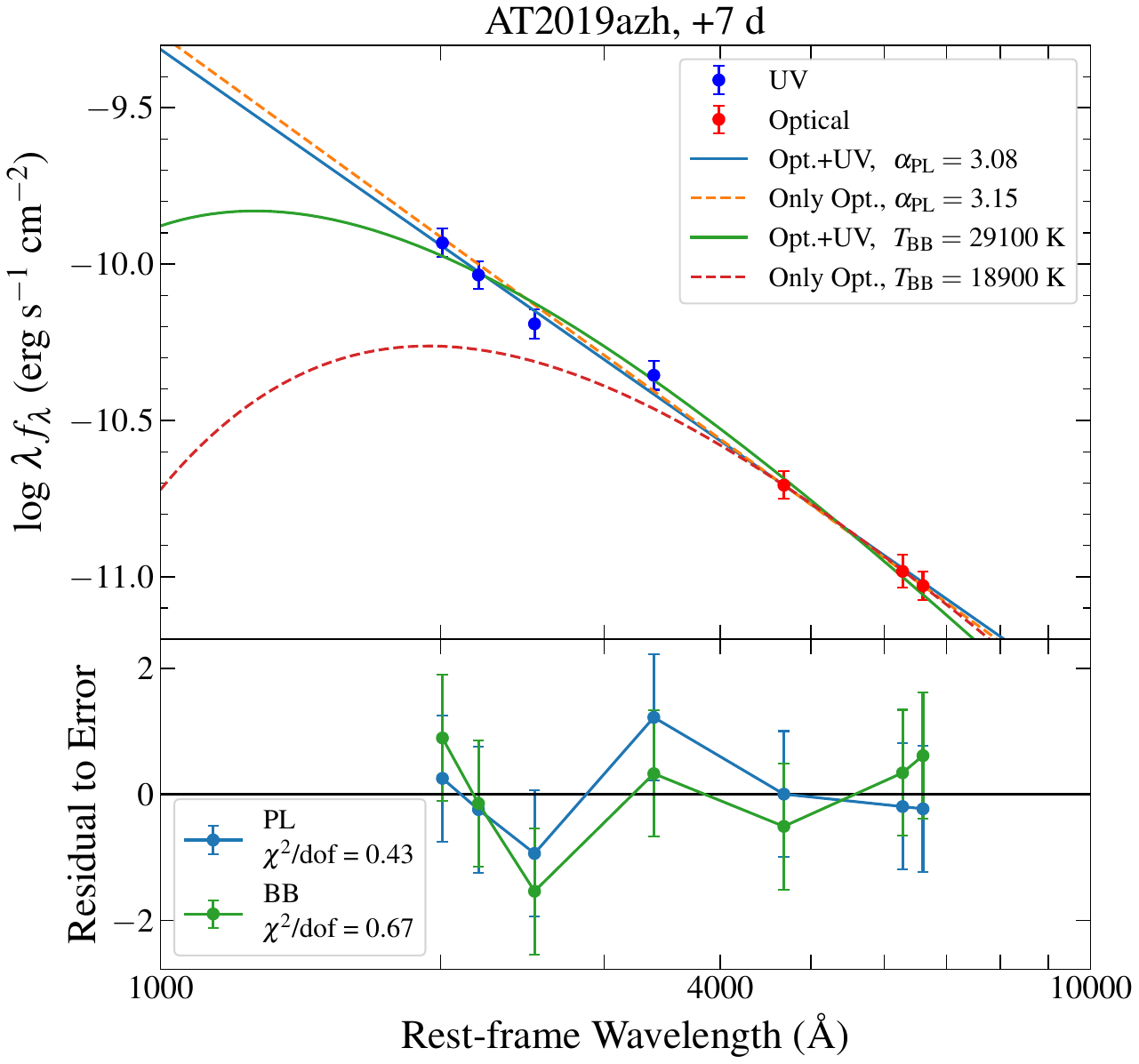}
}
\subfigure[AT 2019dsg]{
	\includegraphics[width=5.5cm]{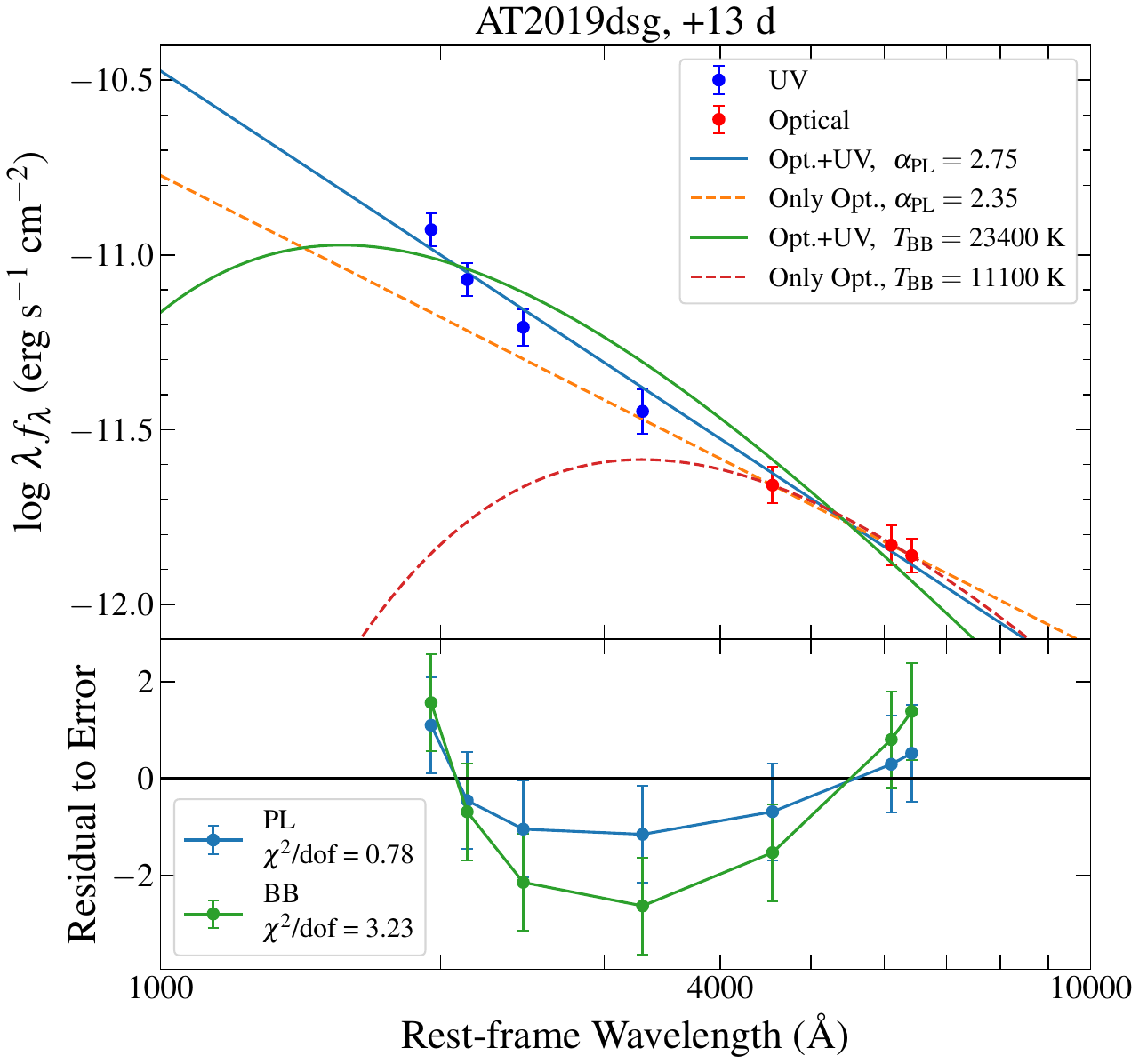}
}
\subfigure[AT 2019qiz]{
	\includegraphics[width=5.5cm]{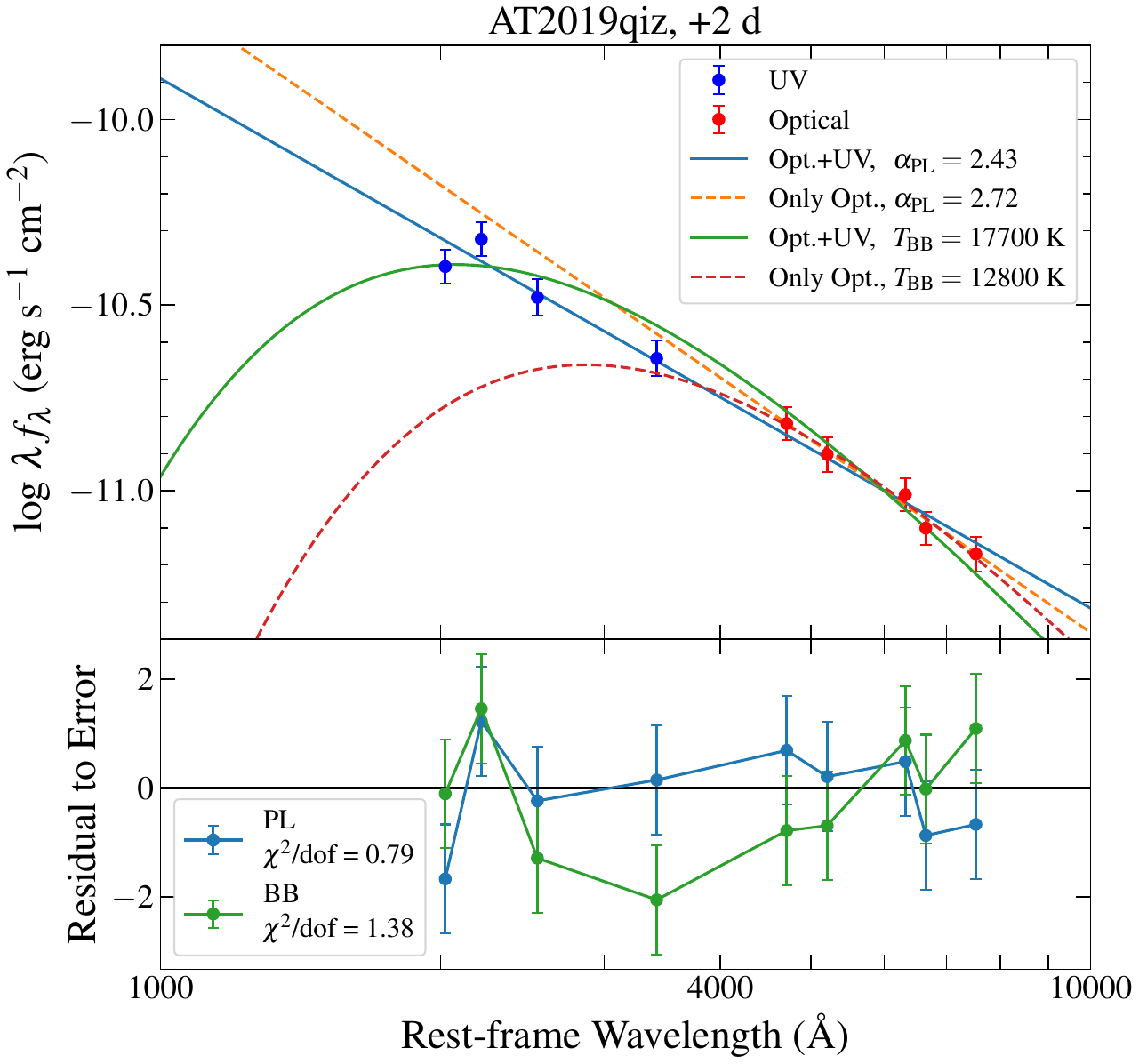}
}
\caption{The rest-frame SEDs around the peak and the best-fit blackbody (BB) and power-law (PL) profiles for AT 2022fpx and four TDEs with high-cadence Swift observations. UV bands are defined as bands with rest-frame wavelength $<$4000 \AA, while others are defined as optical bands. Systematic flux error of 10\% is applied on all measurements. For each subfigure - \textit{Upper panel:} UV and optical bands are plotted in blue and red dots and errors, respectively. The power-law SEDs fitted with both optical and UV bands are drawn in blue lines, while those fitted with only optical bands are plotted in orange dashed lines. The blackbody SEDs fitted with both optical and UV bands are drawn in green lines, while those fitted with only optical bands are plotted in red dashed lines. \textit{Lower panel:} The ratios of residual to error, and the reduced chi-squares.}
\label{fig:compSED}
\end{figure*}

\begin{figure*}[t]
\centering
\subfigure[Blackbody temperature (Optical+UV)]{
	\includegraphics[width=7cm]{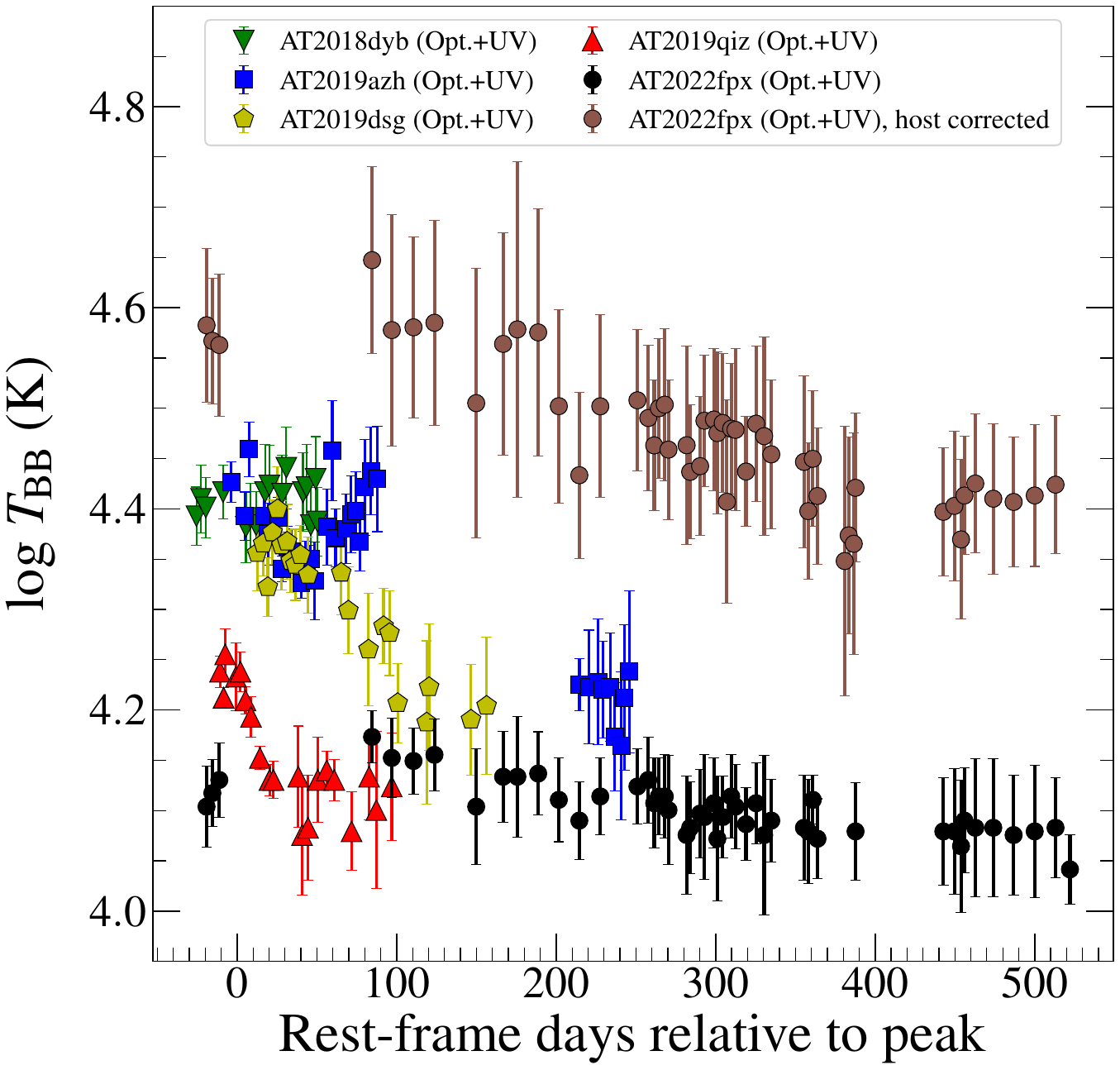}
}
\subfigure[Power-law index (Optical+UV)]{
	\includegraphics[width=7cm]{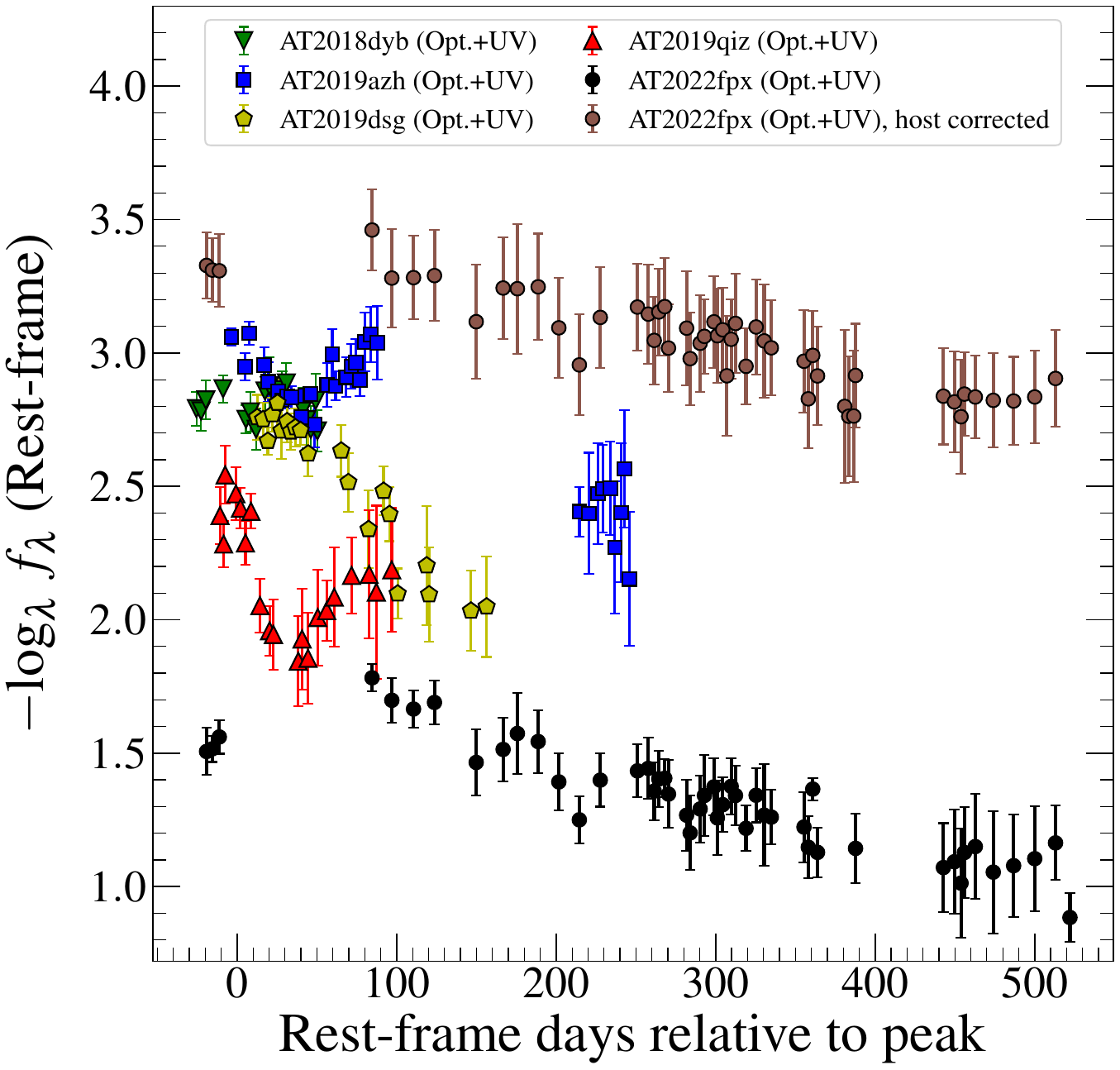}
}
\subfigure[Blackbody temperature (Optical only)]{
	\includegraphics[width=7cm]{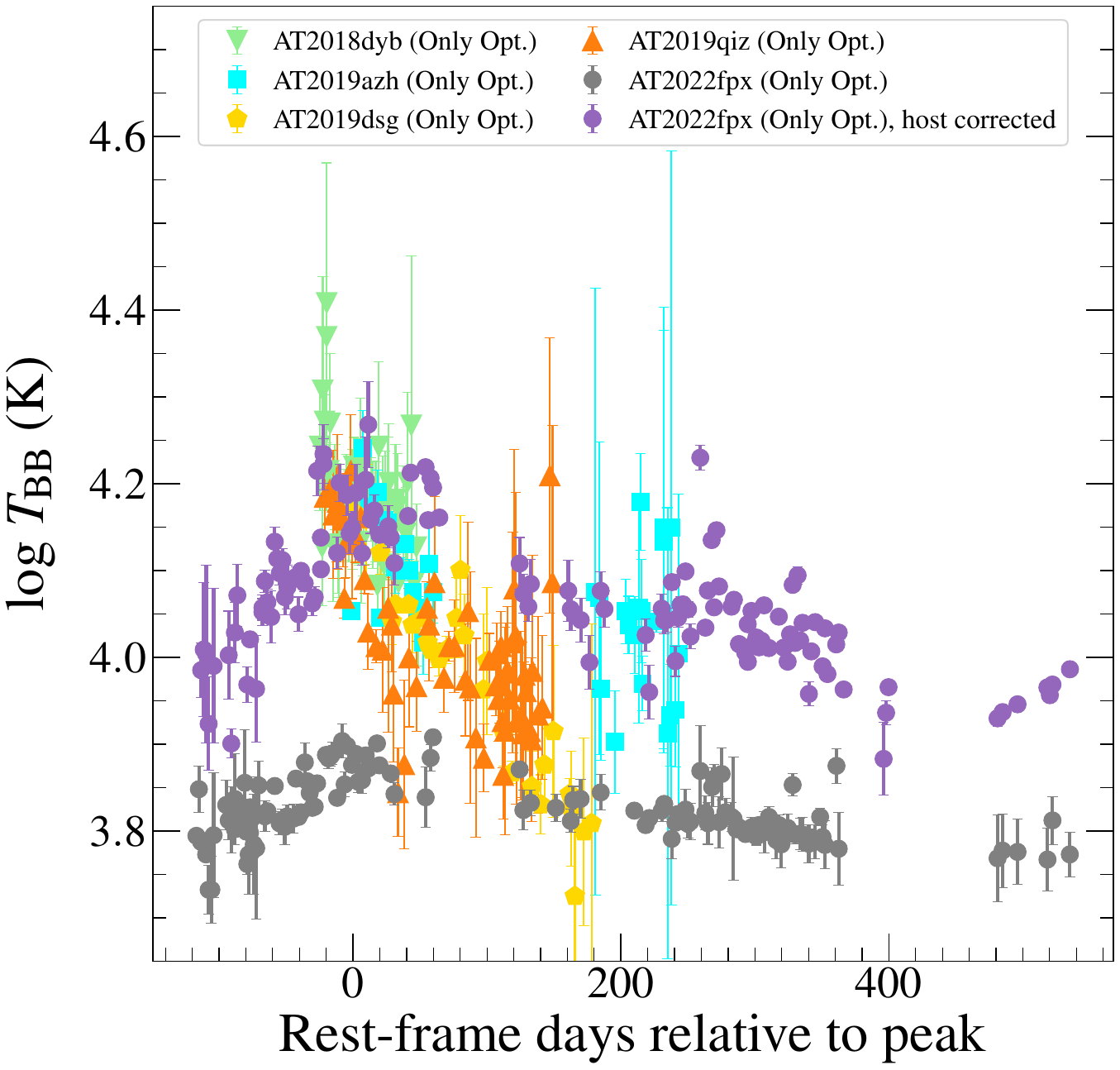}
}
\subfigure[Power-law index (Optical only)]{
	\includegraphics[width=6.75cm]{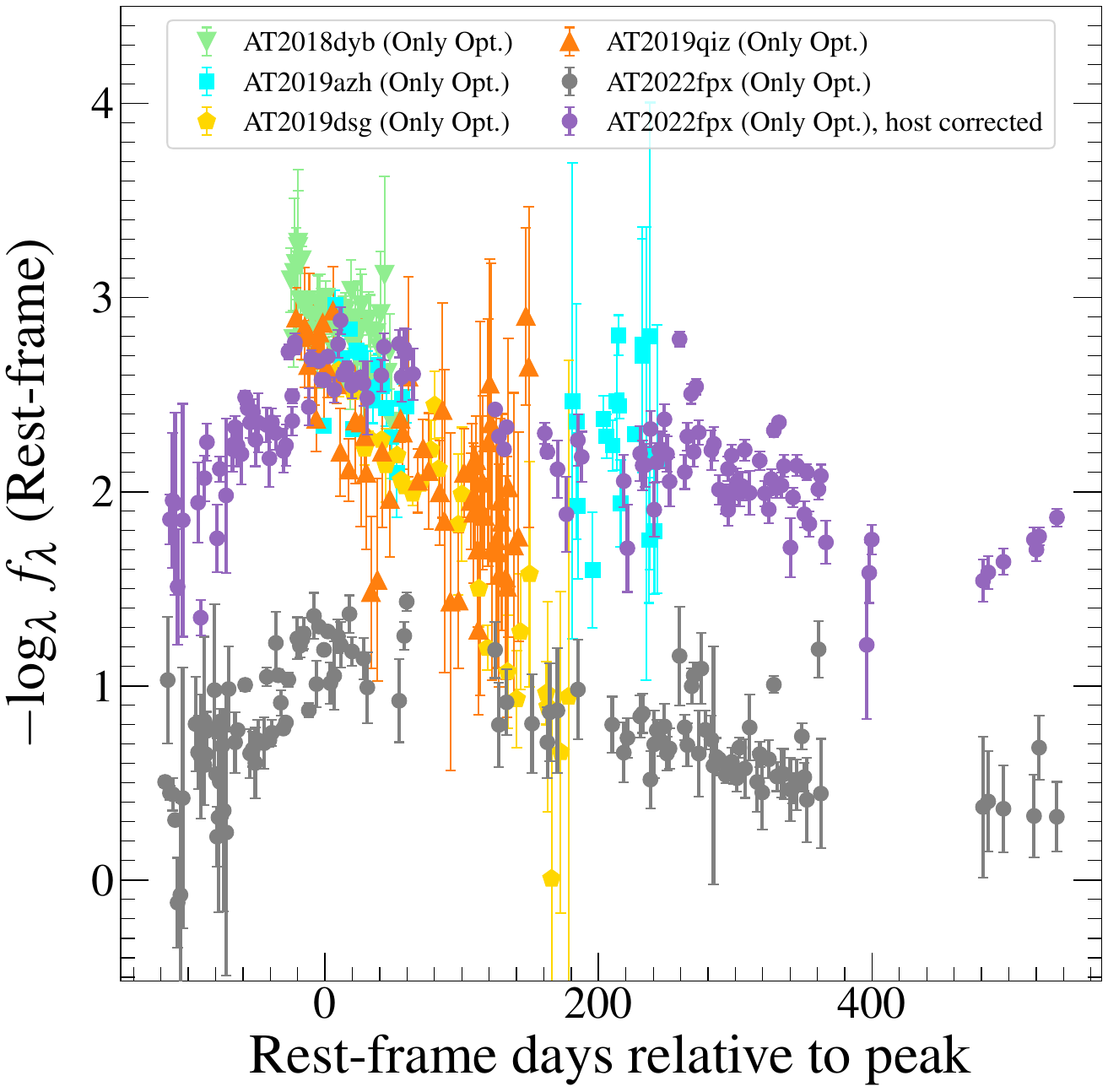}
}
\caption{The temporal evolution of the blackbody temperature and power-law index for AT 2022fpx and four TDEs with high-cadence Swift observations.}
\label{fig:compTDE}
\end{figure*}

We retrieve optical and UV data from these open sources: The ZTF Forced-Photometry Service, the ATLAS forced photometry server and HEASARC (Swift). Then we select four TDEs that have high-cadence Swift observations: AT 2018dyb (ASASSN-18pg), 
AT 2019azh (ASASSN-19dj), AT 2019dsg, AT 2019qiz. 
We note AT 2019ahk (ASASSN-19bt)
was also classified as a TDE and well followed but its host spectrum shows clear Seyfert galaxy features \citep{Holoien2019}, making this classification unreliable. In addition, we retrieve LCO and Swope photometric data ($g$, $r$ and $i$ bands) from the literature \citep{Holoien2020,Hinkle2021,Nicholl2020}. After that, we build the Galactic extinction corrected and host subtracted light curves, and use the same method introduced in Section \ref{sec:flare} to obtain the best-fit blackbody and power-law parameters for each TDE. To provide a clear view of the goodness of the fitting, we fit the rest-frame SEDs near the peaks and plot the best-fit blackbody and power-law fitting results in Figure \ref{fig:compSED}. The temporal evolution is shown in Figure \ref{fig:compTDE}, in which the blackbody temperature and the power-law index are plotted in two subplots, respectively.

From the fitting results, we learn:
\begin{itemize}[noitemsep, topsep=0pt]
    \item If the Balmer decrement is attributed to the dust extinction in the core region, the nature of AT 2022fpx would be an extremely energetic transient with peak optical-UV luminosity of $>$$10^{45}$ erg s$^{-1}$. After correcting this extinction, the optical color $g-r$ turns into blue ($<0$), the blackbody temperature lies in the typical TDE range of $(1-5)\times10^4$ K, and the power-law index is similar with those of typical TDEs. This result suggests that the dust extinction in the core region can consistently explain the red optical color.
    \item The inclusion and exclusion of UV bands in the blackbody fitting cause discrepancy in AT 2022fpx and all four TDEs. The ratio ($T_{\rm bb,opt+UV}/T_{\rm bb,opt}$) for AT 2022fpx is $\sim$209\%; While for four TDEs, the values are $\sim$61\%, $\sim$54\%, $\sim$111\% and $\sim$38\%, respectively. This highlights the importance of UV photometry in the SED fitting of TDEs.
    \item For all four TDEs, their near-peak SEDs near the peak can be well fitted by a power-law model with the index of $\alpha\sim2-3$, and yield a lower reduced chi-square than that of the blackbody model. The main difference between these two models lies in the wavelength range of $<$2000 \AA, since the blackbody flux starts to decline as the wavelength goes down, while the power-law flux continues to rise. As a result, the integrated luminosity is higher when the minimum wavelength is well below 2000 \AA, if a power-law profile is adopted. 
    No significant turning trend displays in all sources, and all power-law index is smaller than the Rayleigh-Jeans limit ($\alpha=4$), these two facts may suggest that the optical-UV SED is part of the
    X-ray-to-IR SED (Figure 5b of \citealt{Dai2018}) that cannot be described by a single blackbody. 
\end{itemize}

\subsection{Summary} \label{sec:summary}
In this section, we discuss the possibility of the SN, AGN and TDE origins, and summarize as follows.
\subsubsection{As a SN}
\noindent \textbf{Optimal solution:} 

AT 2022fpx is originated from the explosion of a SN IIn in a dusty and gas-rich nuclear environment.

\noindent \textbf{Advantages:}
\begin{itemize}[noitemsep, topsep=0pt]
    \item The pure Balmer and helium emission lines agrees with the features of SN IIn.
    \item The broad emission lines with FWHM $\sim$ 2000 km s$^{-1}$ and constantly red optical color have been found in a number of SNe IIn.
\end{itemize}

\noindent \textbf{Disadvantages:}
\begin{itemize}[noitemsep, topsep=0pt]
    \item The ultra-soft X-ray emission ($kT\sim100$ eV) maintained at $>10^{42}$ erg s$^{-1}$ for over two years is unprecedented among SNe IIn.
    \item The extreme iron coronal lines that appear in the rise stage and have constant luminosity of $\sim10^{40}$ erg s$^{-1}$ are also unprecedented among SNe IIn.
    \item The prominent dust echo of $>$$10^{43}$ erg s$^{-1}$ exceeds those of most SNe IIn by one to two orders of magnitude, and comparable with the two brightest dust echoes created by SNe IIn.
    \item The $\sim$120-day rise is one of the longest among SNe IIn.
    \item Nebular lines are absent in the late-time spectrum.
\end{itemize}

\subsubsection{As an AGN}
\noindent\textbf{Optimal solution:} 

AT 2022fpx was a weak AGN before this outburst, and the accretion rate increases sharply after the outburst, resulting in enhanced multiwavelength emission.

\noindent \textbf{Advantages:}
\begin{itemize}[noitemsep, topsep=0pt]
    \item Pre-outburst MIR variability.
    \item Weak AGN contribution in host SED fitting.
    \item The rise timescale is similar as those of outbursts in NLSy1 galaxies, AT 2019brs and AT 2019fdr.
    \item The observed optical-UV SED can be fitted by a power-law profile of $f_{\lambda}\propto\lambda^{-1.58}$. The index 1.58 is close to that of the turn-on AGN iPTF 16bco (1.45) and typical value for type-1 AGN (1.5).
    \item Fe \textsc{ii} continuum existed in the spectrum of the rise stage.
    \item The X-ray luminosity has been maintaining at $\sim$$10^{42}$ erg s$^{-1}$ for $\sim$550 d. Meanwhile, the optical-UV luminosity slowly declines.
\end{itemize}

\noindent \textbf{Disadvantages:}
\begin{itemize}[noitemsep, topsep=0pt]
    \item Whether the outburst and similar outbursts in NLSy1 galaxies are caused by AGN flares remains to be examined.
    \item The Balmer lines have got much weaker recently.
    \item The optical-UV variability is not stochastic, which is different from those of type-1 AGNs. Instead, it monotonously rises and falls slowly like SDSS J1115+0544.
    \item The X-ray spectrum is stably ultra-soft, conform to a power-law profile with $\Gamma=4-5$ or a blackbody model with $kT=100$ eV, which is different from typical AGNs or NLSy1 galaxies but instead similar as TDEs. However, the turn-on AGN candidate SDSS1335+0728 showed a similar X-ray spectrum a few years after the optical outburst. Hence, continuous follow-up X-ray observations should be carried out to track its spectral evolution and determine its origin.
\end{itemize}

\subsubsection{As a TDE}
\noindent\textbf{Optimal solution:} 

AT 2022fpx is a heavily dust-reddened TDE, in which a giant star closely encounters a BH and gets tidally disrupted. An accretion disk quickly forms and generates soft X-ray and EUV photons. These photons are reprocessed to UV, optical and MIR bands by the circumnuclear materials, creating coronal emission lines at the same time. The soft X-ray photons can only be observed after the materials become optically thin. The dust extinction factor $A_V\sim1.14$ mag is estimated by the line ratio of H$\alpha$/H$\beta\sim4$. 

\noindent \textbf{Advantages:}
\begin{itemize}[noitemsep, topsep=0pt]
    \item The ultra-soft and stable X-ray spectrum is well consistent with a TDE.
    \item The relatively long rise timescale can be well explained by a giant-star TDE, under the parameters of $M_\star\sim1.5\,M_\odot$, $R_\star\sim10\,R_\odot$, $\beta\sim2$. While for a main-sequence-star TDE, the rise timescale is nearly irrelevant with the stellar mass and radius, hence a shallow encounter with the BH ($\beta\sim0.7-0.8$) is needed but inconsistent with the released energy. In addition, the evolution also resembles SDSS J1115+0544, which was initially considered as a turn-on AGN but later reclassified as a TDE candidate given the disappearance of broad emission lines 6 years after the optical peak. The fading Balmer emission lines in the recent spectrum appears to support a similar trend. If the broad emission lines indeed disappear in the next few years, it will strongly support the TDE origin. 
    \item After the dust correction, the optical color turns into blue ($g-r<0$), and the SED shape, described by the blackbody temperature or power-law index, is similar as that of typical TDEs (Figure \ref{fig:compSED}).
    \item The MIR echo peaked at $\sim$$10^{43.3}$ erg s$^{-1}$ and faded by only $\sim$0.06 dex in the following $\sim$1.5 yrs. This is well consistent with a heavily dust-attenuated outburst that peaks at $\sim10^{45}$ erg s$^{-1}$.
    \item The extreme iron coronal emission lines are commonly seen among optical TDEs. The Fe \textsc{ii} continuum has also been found in some TDE-in-AGN candidates, such as PS16dtm and PS1-10adi.
\end{itemize}

\noindent \textbf{Disadvantages:}
\begin{itemize}[noitemsep, topsep=0pt]
    \item In order to explain the long rise timescale, unusual stellar profile and impact parameter are introduced. 
    \item The $\sim$550-day long X-ray plateau at $\sim$$10^{42}$ erg s$^{-1}$ contradicts the typical $t^{-5/3}$ decline trend for TDEs, but similar plateaus have been found in some optical-selected TDEs. Regarding that its ultra-soft spectrum is unusual for AGNs, follow-up X-ray observations are still needed to judge its nature.
    \item The intrinsic released energy has exceed $10^{52}$ erg. Considering the extreme slow decline in optical-UV bands, the total released energy might exceed the total energy budget, $\sim0.05(\eta/0.1)M_\star c^2\sim10^{53}(\eta/0.1)$ erg. 
    \item The FWHMs for broad Balmer emission lines are only $\sim$2000 km s$^{-1}$, which is much smaller than those of typical TDEs ($\gtrsim$10000 km s$^{-1}$). However, given the existence of a pre-existing weak AGN and heavy dust attenuation, it is plausible that the broad Balmer emission lines are predominantly produced by the TDE-excited gas in the broad line region of the AGN, which exactly corresponds to the FWHM values. This may be supported by the fading Balmer emission lines indicated by the recent spectrum. 
\end{itemize}

\section{Conclusion} \label{sec:conclusion}
AT 2022fpx is a nuclear transient that is bright in soft X-ray, UV, optical and MIR bands. We investigate its spectral and multiwavelength evolution, as well as its host galaxy. Based on the observational facts, we confirm a pre-existing weak AGN, and conclude that the outburst was very unlikely to be caused by a SN, but equally possible to be caused by a turn-on AGN flare or a heavily dust-attenuated TDE. The potential decisive evidence in the following years could be the disappearance of the broad emission lines (which favors a TDE, slightly supported by the recent fading in the spectrum), the fade of the X-ray/UV/optical emission (which favors a TDE), the hardening and rebrightening of the X-ray emission (which favors a turn-on AGN). Regardless of its origin, we reach several findings during the investigation and comparison with other nuclear transients. 

In Figure \ref{fig:compSED}, we plot the optical-UV SED of AT 2022fpx with those of four typical optical-selected TDEs, and find most of their intrinsic peaks are at $<2000$ \AA. Although the commonly used blackbody model can fit most of these SEDs well, the inclusion or exclusion of UV bands result in largely inconsistent best-fit blackbody temperature and SED shapes. With UV bands, the temperature can be $\sim$40-110\% higher than that without UV bands. The inconsistency could exaggerate if the optical photometry is inaccurate. Most of the optical-UV SEDs of typical TDEs can be better described by a power-law model than the blackbody model, and the inconsistency is also smaller when fitted by a power-law model. The power-law index of 2 to 3 indicates that the optical-UV SED is not part of a single blackbody that peaks at the EUV wavelength.

The optical red color and possible TDE origin of AT 2022fpx are the most intriguing combination, as it challenges the most widely accepted and adopted ``blue color'' criterion for optical TDE selection. Although we still cannot confirm whether it is intrinsic, we however find that this criterion can probably exclude TDEs that have typical optical-UV continuum but either severely contaminated by prominent emission lines (especially H$\alpha$) or heavily dust-reddened. Hence, the potential selection effect may have been imprinted on the whole optical TDE family and remains to be carefully examined.

\begin{acknowledgments}
We appreciate the anonymous referee’s constructive comments, which significantly increase the quality of our manuscript. This work is supported by the National Science Foundation of China (NSFC, grant No. 12192221, 12233008), the National Key R\&D Program of China (2023YFA1608100), the Strategic Priority Research Program of the Chinese Academy of Sciences (grant No. XDB0550200, XDB0550202), the Cyrus Chun Ying Tang Foundations, and the 111 Project for “Observational and Theoretical Research on Dark Matter and Dark Energy” (B23042).
We acknowledge the support of the staff of the Lijiang 2.4m telescope. Funding for the telescope has been provided by Chinese Academy of Sciences and the People's Government of Yunnan Province. We thank the Swift Science Operations team for accepting our ToO requests and arranging the observations. We thank Muryel Guolo and Karri Koljonen for submitting Swift ToO requests. We thank Jane Dai for the useful discussion. Z.Y.L. sincerely thanks the UK Swift Science Data Centre (UKSSDC) helpdesk (especially Phil and Kim) for the kind instructions and \textit{Swift} replies on the reduction of XRT data, and thanks Robert Wiegand for the help on the reduction of UVOT data. Z.Y.L. thanks Mederic Boquien for solving questions on the \texttt{CIGALE} package. The ZTF forced-photometry service was funded under the Heising-Simons Foundation grant \#12540303 (PI: Graham). This research
uses data obtained through the Telescope Access Program
(TAP). Observations with the Hale Telescope at Palomar Observatory were obtained as part of an agreement between the
National Astronomical Observatories, Chinese Academy of
Sciences, and the California Institute of Technology. This
work is based on the data obtained with Einstein Probe, a space mission supported by Strategic Priority Program on Space Science of Chinese Academy of Sciences, in collaboration with ESA, MPE and CNES (Grant No. XDA15310000).
\end{acknowledgments}

\begin{contribution}

ZYL conducts the Swift and EP follow-up observations, reduces their data and all archival data, and writes this manuscript. NJ and XK provides crucial insights to the scientific meaning of this source, and also the co-PIs for the EP proposal. YW conducts most of the spectroscopic follow-up observations and data reduction. SH provides instructions on the X-ray data reduction and important viewpoint on the TDE origin. ZSL offers key instructions in calculating the dust correction factor. CQ offers help in obtaining the latest spectrum. TX offers help in writing the EP proposal and obtaining EP data.



\end{contribution}

%



\appendix

\section{Detailed information for transients used for comparison}\label{sec:detinfo}
Transients used for comparison in Section \ref{sec:trans4comp} are grouped by type and introduced below:
\vspace{10pt}

\noindent1. TDE (candidate):

\textbf{AT 2019qiz.} AT 2019qiz is a typical TDE at $z=0.0151$, whose spectrum displays broad ($\gtrsim$$10000$ km s$^{-1}$) Balmer, helium and Bowen fluorescence lines above the blue continuum \citep{Nicholl2020}. 

\textbf{SDSS J1115+0544.} SDSS J1115 was classified as a turn-on AGN triggered by an outburst \citep{Yan2019}. Its pre-outburst SDSS spectrum is consistent with a quiescent early-type galaxy at $z=0.0899$. After the outburst, broad H$\alpha$ and H$\beta$ lines with FWHM $\sim3750$ km s$^{-1}$ and narrow coronal [Fe \textsc{vii}] lines exhibited in the spectrum. The optical light curve strikingly resembles that of AT 2022fpx. It rose to the peak of $M_V\approx-19.1$ in $\sim120$ days, and dropped to a plateau in $\sim200$ days. After that, its luminosity maintained at the plateau for $\sim600$ days. A similar dust echo of $L_{\rm MIR}\approx2\times10^{43}$ erg s$^{-1}$ and pre-flare color of W1$-$W2 $=0.173$ was unveiled by WISE. The X-ray emission was not detected, yielding a 90\% upper limit of $\sim1.4\times 10^{42}$ erg s$^{-1}$, which is different from the delayed X-ray emission of AT 2022fpx. However, \citet{Wang2022a} reported that the broad emission lines surprisingly disappeared $\sim$6 years after the optical peak. Based on this disappearance, it was reclassified as a TDE candidate. The light curves reveal a slow decline trend in the luminosity after the plateau, which is also against a turn-on AGN.

\textbf{SDSS J0952+2143.} SDSS J0952 is the best-studied extreme coronal line emitter (ECLE) at $z=0.0789$. It had undergone an outburst $\sim600$ days before the SDSS spectrum was taken. The SDSS spectrum shows prominent iron coronal emission lines with luminosity of $\sim10^{40}$ erg s$^{-1}$ and FWHM of a few $\times10^2$ km s$^{-1}$. In the next few years, the continuum, iron coronal emission lines and optical light curves continuously faded \citep{Komossa2008,Komossa2009,Wang2011,Yang2013,Palaversa2016}. These features and their evolution are fairly consistent with AT 2022fpx, except that AT 2022fpx has not shown the fading trend of iron coronal emission lines. Its TDE origin was first proposed by \citet{Komossa2008} and confirmed by \citet{Palaversa2016}.

\textbf{AT 2018dyk.} Before the outburst, the host galaxy was a LINER. In 2018, an outburst occurred in X-ray, UV, optical and MIR wavelengths. The peak of the X-ray luminosity lags behind that of optical by $\sim$140 d. The X-ray spectrum is soft and conform to a power-law profile of $\Gamma=3$. The source was initially classified as a changing-look AGN \citep{Frederick2019}, but later revisited and reclassified as a TDE \citep{Huang2023,Clark2025}. It is also an ECLE and Fe \textsc{ii} continuum emitter. 

\textbf{PS16dtm.} First discovered by \citet{Blanchard2017} as a TDE in a NLSy1 galaxy. After a strong outburst in optical and MIR bands, the pre-existing X-ray emission has been suppressed until now. Prominent Fe \textsc{ii} emission lines exhibit in rest-frame wavelength of $\sim$4500 \AA\ and $\sim$5300 \AA. Its MIR echo has lasted for more than 8 yrs and is still maintaining at $>10^{43}$ erg s$^{-1}$, which is believed to be supported by an extraordinarily energetic but heavily dust-attenuated nuclear outburst ($>10^{46}$ erg s$^{-1}$) \citep{Jiang2025}.

\vspace{10pt}

\noindent2. SN:

\textbf{SN2008iy.} AT 2022fpx displayed a $\sim$120-day rise stage, which would the second longest if it is a SN IIn (See Section \ref{sec:onSN}). The longest record is crazy: $\sim$400 days, held by SN2008iy at $z=0.0411$ \citep{Miller2010}. 

\textbf{SN2010jl.} SN2010jl is a SN IIn close to the center of galaxy UGC 5189A at $z=0.01062$. It displayed a $\sim200$-day plateau in optical bands after the peak. At late time, it exhibited a IR dust echo of $L_{\rm IR}\gtrsim5\times10^{42}$ erg s$^{-1}$ and a hard X-ray spectrum with unabsorbed $0.2-10$ keV luminosity of $L_{\rm X}\sim7\times10^{41}$ erg s$^{-1}$ \citep{Zhang2012,Chandra2012,Fransson2014}. These features resemble AT 2022fpx, except that the X-ray spectrum is much harder.

\vspace{10pt}


\vspace{10pt}

\noindent3. Turn-on / Flaring AGN:

\textbf{iPTF 16bco.} iPTF 16bco was reported as a turn-on quasar by \citet{Gezari2017a}. Before the ``turn-on", the host galaxy was a LINER at $z=0.2368$. The quasar continuum increased by a factor of $>$10 on a timescale of $\lesssim$1 year in the quasar rest frame. The continuum of the difference spectrum can be fitted with a power-law of $f_{\lambda}\propto\lambda^{-1.45}$. After the ``turn-on'', the optical luminosity has maintained at the high level, and shows stochastic variability that is similar with that of common type-1 AGNs. Its X-ray spectrum can be fitted with an absorbed power-law with $\Gamma=2.1\pm0.5$, which is typical for a type-1 AGN, $\Gamma=1.75$ \citep{Ricci2017}. 

\textbf{AT 2019brs.} Before the outburst, the host galaxy was a narrow-line Seyfert 1 (NLSy1) galaxy. A prominent outburst occurred in 2019 in UV, optical and MIR bands. Its rest-frame rise timescale is $\sim$120 d, similar to that of AT 2022fpx, while its black hole mass is estimated to be $M_{\rm BH}\sim10^{7.2-8.2}\,M_{\odot}$, which is $\sim$10$-$100 times higher than the estimated $M_{\rm BH}$ of AT 2022fpx. It is relatively quiet in X-ray, as only one XRT epoch can marginally detected its X-ray emission throughout the optical-bright stage. The outburst was classified as an AGN flare \citep{Frederick2021}.

\textbf{AT 2019fdr.} Before the outburst, the host galaxy was also a NLSy1 galaxy. A prominent outburst occurred in 2019 in UV, optical and MIR bands. During the subsequent declining phase, it got redder in the optical color, and the luminosity once reached a plateau and maintained for $\sim$120 rest-frame days. It is quiet in X-ray, as no X-ray counterpart has been found. It has been classified as a TDE \citep{Frederick2021} and a SLSN-II \citep{Yan2019}.

\textbf{SDSS1335+0728.} SDSS1335+0728 was reported as a turning-on AGN by \citet{Sanchez2024}. Before the ``turn-on", the host galaxy was a star-forming galaxy at $z=0.024$. After the ``turn-on'', the optical luminosity has maintained at the high level, and shows stochastic variability that is similar with that of common type-1 AGNs. Unlike iPTF 16bco, its X-ray spectrum is extremely soft, which can be described by a blackbody model of $kT=80\pm7$ eV. Nonetheless, the authors note that an exotic TDE still cannot be excluded.

\bibliography{sample701}{}
\bibliographystyle{aasjournalv7}



\end{document}